\documentclass[onecolumn,11pt,msmath,amsmath,amssymb,aps]{revtex4-2}
  \DeclareUnicodeCharacter{2212}{-}
\DeclareUnicodeCharacter{2009}{-}
\usepackage[colorlinks=true,urlcolor=blue,linkcolor=blue,citecolor=blue]{hyperref}
\usepackage{subfigure}
\usepackage{tablefootnote}
\usepackage{wrapfig}
\usepackage{graphicx}
\usepackage{textcomp}
\usepackage{bm}
\usepackage{xfrac}
\usepackage{color}
\usepackage[dvipsnames]{xcolor}
\usepackage{setspace}
\usepackage{silence}
\usepackage{epstopdf}
\RequirePackage[normalem]{ulem} 
\RequirePackage{color}
\definecolor{RED}{rgb}{1,0,0}
\definecolor{BLUE}{rgb}{0,0,1}

\makeatletter
    \renewcommand\@make@capt@title[2]{%
     \@ifx@empty\float@link{\@firstofone}{\expandafter\href\expandafter{\float@link}}%
      {\textbf{#1}}\@caption@fignum@sep#2\quad}%
    \makeatother
\usepackage{array}
\newcolumntype{C}[1]{>{\centering\arraybackslash}m{#1}}

\begin{document}

\title{Quantum Transport and Potential of Topological States for Thermoelectricity in Bi$_2$Te$_3$ Thin Films}

\author{Prosper Ngabonziza}
\email{p.ngabonziza@fkf.mpg.de}
\affiliation{Max Planck Institute for Solid State Research, \\ 
Heisenbergstraße 1, 70569 Stuttgart, Germany}
\affiliation{Department of Physics, University of Johannesburg, \\
P.O. Box 524, Auckland Park 2006, Johannesburg, South Africa}

\date{\today}
\begin{abstract}
This paper reviews recent developments in quantum transport and it presents recent efforts to explore the contribution of topological insulator boundary states to thermoelectricity in Bi$_2$Te$_3$ thin films. Although Bi$_2$Te$_3$ has been used as a thermoelectric material for many years, it is only recently that thin films of this material have been synthesized as 3D topological insulators with interesting physics and potential applications related to topologically protected surface states. A major bottleneck in Bi$_2$Te$_3$ thin films has been eliminating its bulk conductivity while increasing its crystal quality. The ability to grow epitaxial films with high crystal quality and to fabricate sophisticated Bi$_2$Te$_3$-based devices is attractive for implementing a variety of topological quantum devices and  exploring the potential of topological states to improve thermoelectric properties.  Special emphasis is laid on preparing low-defect-density Bi$_2$Te$_3$ epitaxial films, gate-tuning of normal-state transport and Josephson supercurrent in topological insulator/superconductor hybrid devices. Prospective quantum transport experiments on Bi$_2$Te$_3$ thin-film devices are discussed as well. Finally, an overview of current progress on
the contribution of topological insulator boundary states to thermoelectricity is presented. Future explorations to reveal the potential of topological states for improving thermoelectric properties of  Bi$_2$Te$_3$ films and realizing high-performance thermoelectric devices are discussed.
\vfill
\end{abstract}

\newpage
\maketitle
\section*{\Large{1. I\lowercase{ntroduction}}}

A topological insulator (TI) is a material that has an insulating interior (bulk) and conducting  boundaries. The bulk of a TI is characterized by an energy gap in the electronic band structure, whereas the edges or surfaces host non-gapped states crossing the bulk band gap, thus enabling conduction on the boundaries of the material~\cite{CLKane_2005}. The surface states of a TI mimic relativistic Dirac electrons because of their linear energy-momentum relation~\cite{LFu_2007}. The appearance of metallic surface/edge states is due to the strong spin-orbit coupling in the material, which causes a band inversion. 
TIs have generated much interest, not only as fundamental new electronic states of matter but also for enabling the creation of devices that exploit the unusual properties of topological surface/edge states. This allows  future technological applications in such fields as spintronics, topological quantum computation and others~\cite{LFu_2008,ARAkhmerov_2009,DHsieh_2009,CJozwiak_2007,IAppelbauma_2011,MHe_2019}.

	TIs were first predicted theoretically in 2D quantum-spin Hall  systems (HgTe/CdTe)~\cite{BABernevig_2006,BABernevig_2006-01,CLKane_2005}. Shortly thereafter, the first 3D TI was experimentally realized in Bi$_x$Sb$_{1-x}$ alloys~\cite{DHsieh_2008,DHsieh_2009-01}. Soon afterwards, the materials Sb$_2$Te$_3$, Bi$_2$Te$_3$, and Bi$_2$Se$_3$ were predicted and experimentally confirmed to be 3D TIs. 
These latter compounds (Sb$_2$Te$_3$, Bi$_2$Te$_3$, and Bi$_2$Se$_3$) have the advantage of exhibiting surface states with a single Dirac cone~\cite{HZhang_2009,YXia_2009,YLChen_2009}, which makes them candidate materials for the exploration of exotic topological states such as Majorana bound states and magnetic monopoles in condensed-matter systems~\cite{LFu_2008,XLQi_2009}. The confirmation of the existence of 3D TIs has led to an explosion of research in this field, both theoretically and experimentally. Since then, many more materials have been added to the list of 3D TIs, and their diverse potential technological applications have been explored~\cite{YAndo_2013,MZHasan_2013,MHe_2019}. 

This review focuses on Bi$_2$Te$_3$ epitaxial films. The material Bi$_2$Te$_3$ has a rhombohedral crystal structure consisting of quintuple layers (QLs) of bismuth and tellurium atoms bound to each other (stacking sequence: Te[1]-Bi-Te[2]-Bi-Te[1]) through weak Van der Waals (VdW) forces, see Figs.~\ref{fig:Fig-01}\textcolor{blue}{(a)} and~\textcolor{blue}{(b)}. Bi$_2$Te$_3$ is a prime member of 3D TIs that is most extensively studied due not only to its fascinating TI properties~\cite{YLChen_2009,ZAlpichshev_2010,YLi_2010,PNgabonziza_2016,VMPereira_2021,PNgabonziza_situ_2018}, but also its excellent room-temperature thermoelectric (TE) properties~\cite{GJSnyder_2008,CUher_2016}. By exploiting these properties, Bi$_2$Te$_3$ has been used to fabricate flexible TE devices with a wide range of technological applications, for example in self-powered wearable electronics, electrical power-generation devices and precision temperature control of microchips~\cite{RVenkatasubramanian_2001,IChowdhury_2001,LWang_2018,QJin_2018,IPetsagkourakis_2018,CDagdeviren_2017,ITWitting_2019}. Topological surface states and TE properties of  Bi$_2$Te$_3$ are intimately related to the position of the Fermi level $(E_\textrm{F})$ in the electronic band structure~\cite{GWang_2011,YLi_2010,PNgabonziza_2015,KHoefer_2014} and the concentration of intrinsic defects~\cite{CTang_2019,TZhu_2016,LHu_2014,NPeranio_2013,PNgabonziza_2018,KShrestha_2017}. 
Furthermore, interest in Bi$_2$Te$_3$  has been  heightened recently by proposals for topological surface-state-enhanced TE devices~\cite{PGhaemi_2010,OATretiakov_2011,ZFan_2012,NXu_2017}, which underscore the need for systematic experimental explorations of the contribution of TI boundary states to thermoelectricity. 

However, as bulk conduction in Bi$_2$Te$_3$ complicates the direct exploitation of TI surface effects, great care must be taken to improve the crystal quality and obtain samples with fewer intrinsic defects. The most promising approach to improve the crystal quality and obtain samples with minimal intrinsic defects and high surface mobilities is to prepare high-quality intrinsically insulating Bi$_2$Te$_3$ thin films~\cite{YLi_2010,PNgabonziza_2015,KHoefer_2014,PNgabonziza_2016}. This paper reviews the latest progress in studying the quantum transport properties of Bi$_2$Te$_3$ epitaxial films and    the contribution of TI boundary states to thermoelectricity. Recent advances in thin-film growth, magnetotransport and induced superconducting proximity effect in Bi$_2$Te$_3$ thin films are discussed. Finally, an overview of current efforts to explore the contribution of topological boundary states to thermoelectricity in epitaxial Bi$_2$Te$_3$ thin films  is presented.

\paragraph*{}
\section*{\Large{2. O\lowercase{ptimizing the} G\lowercase{rowth} P\lowercase{arameters of} B\lowercase{i$_2$}T\lowercase{e$_3$} t\lowercase{hin} f\lowercase{ilms}}}
\paragraph*{} 
To study the salient topological surface states and TE properties of Bi$_2$Te$_3$, thin films of this material have been synthesized over the past few years using different experimental  methods. The most widely used techniques include, but are not limited to: molecular beam epitaxy (MBE) \cite{LHe_2013,YLi_2010,GWang_2011,PNgabonziza_2015,NPeranio_2013}, pulsed laser deposition (PLD) \cite{HObara_2009,Zhang_2012,LZhaoliang_2020}, metal-organic chemical vapor deposition (MOCVD) \cite{HWYou_2012,CHelin_2012,KCKim_2020}, hot-wall epitaxy (HWE) \cite{MFerhat_2000,YTakagaki_2012,SCharar_2000}, electrodeposition~\cite{AZimmer_2008,SLi_2008,AZimmer_2007}, ion beam sputtering deposition~\cite{ZHZheng_2012,ZHZheng_2010,ZHZheng_2014}, and magnetron sputtering deposition~\cite{MGoto_2017,YSasaki_2014,ZKCai_2013}. Amongst these techniques, MBE has been especially well established for growing high-quality crystalline thin films with precise control of film thickness down to atomic layers~\cite{LHe_2013}.
In addition, MBE makes it easy to combine different TI compounds, for example to form $p$\nobreakdash-$n$ junctions~\cite{TMayer_2021,MLanius_2016,MEschbach_2015} or TIs with other materials such as TI-ferromagnetic insulator heterostructures~\cite{SJChang_2018,VMPereira_2020,VMPereira_2021}. In particular, intrinsically  insulating Bi$_2$Te$_3$ samples with low bulk-defect density have been prepared by carefully controlling the MBE growth conditions~\cite{YLi_2010,PNgabonziza_2015,KHoefer_2014,KHoefer_2015}. Such samples allow a fine-tuning of  $E_\textrm{F}$ in the band gap, which is of key importance for harnessing special properties of topological surface states in order to optimize the TE performance of TI nanostructures~\cite{NXu_2017}. The high flexibility in combining different elements, the good control over  film thickness and the possibility of uniform deposition over a large area make MBE the most versatile method for  synthesizing multifunctional TI/TE Bi$_2$Te$_3$ nanomaterials. This section focuses on recent experimental endeavors to optimize  MBE growth conditions for achieving high-quality Bi$_2$Te$_3$ with low defect density.

\begin{figure*}[t]
\centering
\includegraphics[width=1\textwidth]{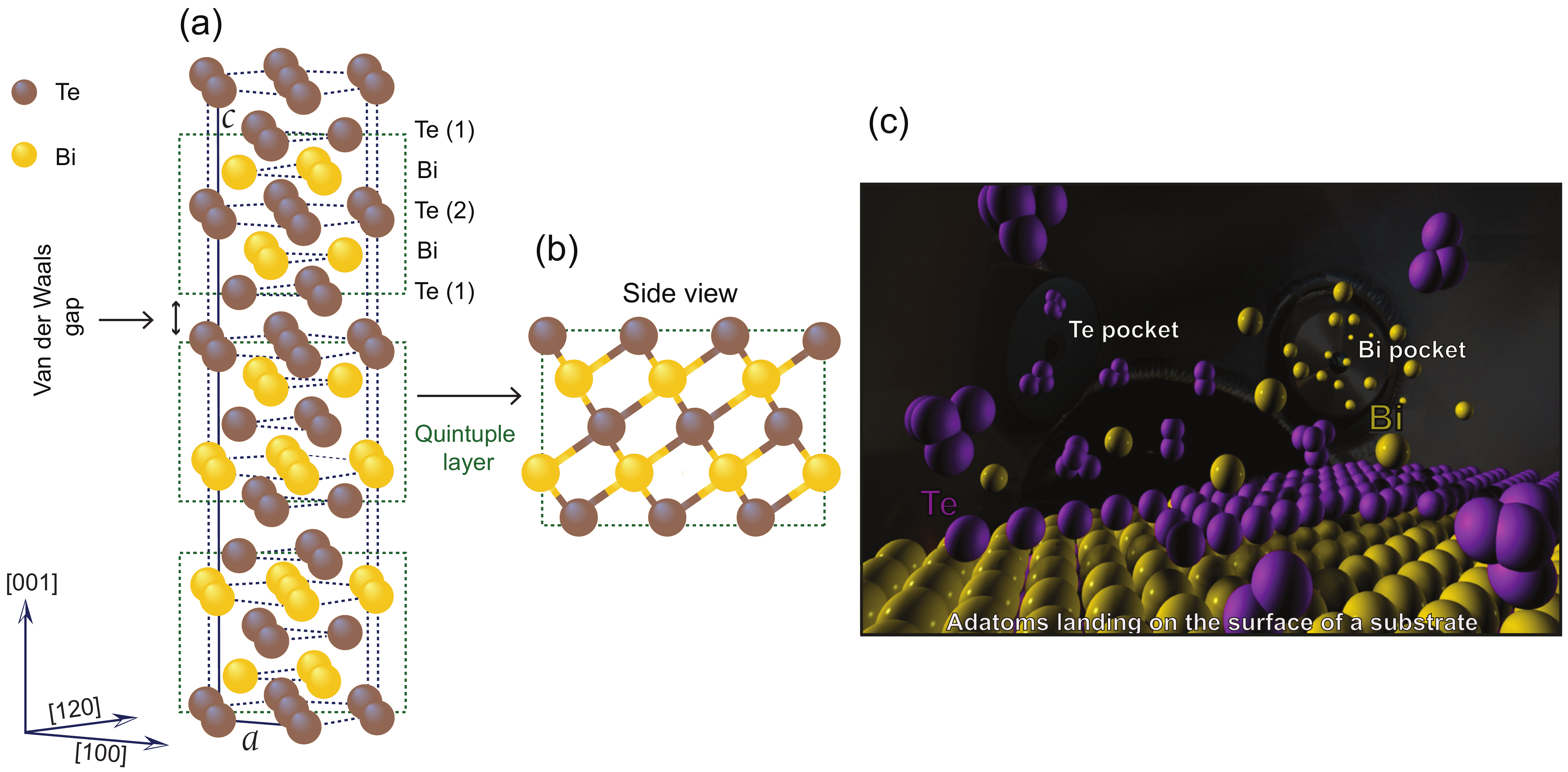}
\caption{\linespread{1.5} (a) Crystalline structure of Bi$_2$Te$_3$. The unit cell is formed by  stacking  three QLs in [001] direction. VdW  forces between adjacent QLs through Te atoms is much weaker than that between Te and Bi inside the QL.  (b)~Side view of one QL slab of Bi$_2$Te$_3$. (c)~Te-rich growth environment in the MBE growth process, which is the ideal condition for obtaining high-quality Bi$_2$Te$_3$ TI thin films. (c)~Reproduced and adapted with permission from~\cite{GWang_2013},  \copyright ~2012 Elsevier Science, all rights reserved.}
 \label{fig:Fig-01}
\end{figure*}

During MBE growth, elemental or compound materials of interest are generally evaporated from sources contained in electrically heated furnaces or crucibles~\cite{BAJoyce_1985,MHenini_2013}. The evaporated species pass through an ultra-high vacuum (UHV) environment and are deposited on a substrate that is hundreds of millimeters away from the evaporated materials. Under such UHV conditions, the incorporation of contaminants from the background gas is minimized. The temperature of the sources is set according to the type of material used, because different materials have different melting points and vapor pressures. The MBE growth process of Bi$_2$Te$_3$, and for other 3D TIs in general, is mainly affected by the atomic flux ratio of the elements, the UHV conditions in the growth environment and the substrate temperature~\cite{LHe_2013,MHenini_2013}. 

To maximize the surface diffusion for confining nucleation and growth at the proper atomic sites, the substrate temperature is often tuned to an optimum value. Compared to bulk crystals (typically synthesized at temperatures above 600\textdegree  C)~\cite{RSultana_2018,CTang_2019}, high-quality thin films of Bi$_2$Te$_3$ are grown by MBE at much lower temperatures in the range of 200–350\textdegree C~\cite{LHe_2013,PNgabonziza_2015,SEHarrison_2013,KHoefer_2014}. The reduced kinetics at low growth temperatures helps  minimize the formation of Te vacancies in MBE-grown thin films. Furthermore, Bi$_2$Te$_3$ bulk crystals can be grown in both $n$- and $p$-types~\cite{DXQu_2010,YAndo_2013}. This is because the Te anti-site defects Te$^{+}_{\text{Bi}}$ (Te ion sitting on a Bi lattice site) or Bi anti-site defects Bi$^{-}_{\text{Te}}$ (Bi ion sitting on a Te lattice site) are formed in Te-rich or Bi-rich conditions, respectively~\cite{YAndo_2013}. Te$^{+}_{\text{Bi}}$ is a donor, whereas Bi$^{-}_{\text{Te}}$ is an acceptor, thus suggesting that, by carefully controlling the MBE growth conditions, it would be possible to prepare high-quality bulk insulating Bi$_2$Te$_3$ thin films. 

Figure \ref{fig:Fig-01}\textcolor{blue}{(c)} illustrates the Te-rich growth dynamics required for preparing high-quality Bi$_2$Te$_3$ films using MBE~\cite{LHe_2013,JKrumrain_2011}. To obtain high-quality films, it is necessary to control the atomic flux ratio of the Bi and Te elements~\cite{LHe_2013,YLiu_2013,YLi_2010}. In  practice, to prepare Bi$_2$Te$_3$, the growth should be performed in the tellurium-rich side of the phase diagram in order to avoid Te vacancies. This Te-rich condition helps  achieve the correct stoichiometry (a ratio of Te:Bi = 3:2)  to ensure the best film growth. To obtain stoichiometric Bi$_2$Te$_3$ thin films, it is often suggested that the Te/Bi atomic ratio should be in the range of 8 to 20~\cite{LHe_2013,YLi_2010}. Furthermore, when using thermal evaporation, the condition $T_{\text{Bi}}>T_{\text{sub}} >T_{\text{Te}}$ (where $T_{\text{Bi}}$, $T_{\text{sub}}$ and $T_{\text{Te}}$ are the Bi-cell, substrate and Te-cell temperatures, respectively) should be satisfied. 

\begin{flushleft} 
\textit{2.1 Surface Passivation and Surface Morphology}
\end{flushleft}
\paragraph*{} 
The lattice constant mismatch between a substrate and a film is a major challenge for achieving epitaxial growth of thin films. A large lattice mismatch between the constituent materials leads to defects and worsens the crystal quality of the epitaxially grown film. However, for growing Bi$_2$Te$_3$ films (also for other 3D TI systems), the lattice-match criteria 
are drastically relaxed because the unit cells are coupled by VdW interactions, see Figs.~\ref{fig:Fig-01}\textcolor{blue}{(a)} and \textcolor{blue}{(b)}. This growth preference is known as VdW epitaxy~\cite{AKoma_1992,AKoma_1999}. Therefore, a variety of substrates can be chosen, and relatively good Bi$_2$Te$_3$ films have been obtained for each. Substrates reported so far include, but are not limited to: Si [111]~\cite{AFulop_2014,HWLiu_2010}, Al$_2$O$_3$ [0001]~\cite{SEHarrison_2013,PNgabonziza_2015}, SrTiO$_3$ [111]~\cite{PNgabonziza_2015,PNgabonziza_2016}, GaAs [111]B~\cite{ZZeng_2013,XYu_2012}, GaAs [001]~\cite{XLiu_2011,XLiu_2012}, BaF$_2$ [111]~\cite{HSteiner_2014,OCaha_2013} and Mica~\cite{KWang_2013,RRapacz_2015}. For purposes of avoiding the contribution of conduction from the substrate in electronic transport measurements, insulating substrates are preferred. In particular, insulating oxide substrates with a high relative dielectric constant at low temperatures offer the opportunity of gate-tuning the film's transport properties using the substrates as a back gate~\cite{PNgabonziza_2016,XHe_2013}. Although the atomically well-defined single-termination of these substrates is not a requirement, proper surface preparation is desirable to obtain good templates for epitaxial growth~\cite{ABiswas_2017,ANonoTchiomo_2018,WBraun_2020,JChang_2008,FSanchez_2014}.

\begin{figure*}[!t]
\centering
\includegraphics[width=1\textwidth]{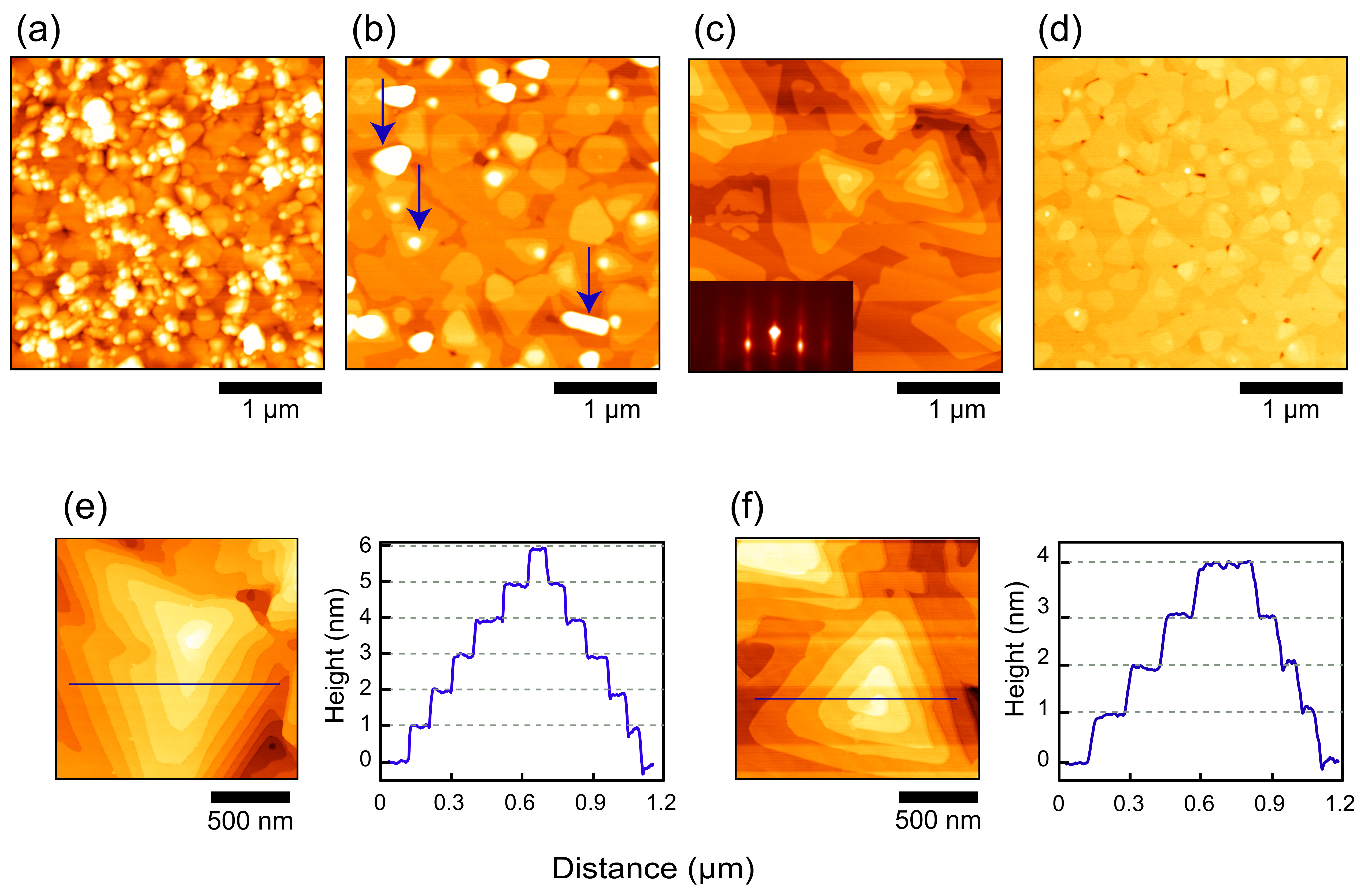}
\caption{\linespread{1.5}Growth morphology of Bi$_2$Te$_3$ films of different thicknesses grown on distinct substrates at the same optimal substrate temperature of the nucleation layer (170\textdegree C), but with different temperatures for the second layer (170, 210 and 250\textdegree C). Surface morphology of 15-nm Bi$_2$Te$_3$ films grown on (111)-oriented SrTiO$_3$ (STO) at different substrate temperatures (a) 170\textdegree C, (b) 210\textdegree C, and (c)--(d) 250\textdegree C. The optimized growth temperature (for the second layer) was found to be around 250\textdegree C. The sample in (c) was annealed for 5 min, whereas the sample in (d) was annealed for 30 min  immediately after growth. The three arrows in (b) point to small and larger 3D defects and quasi-1D nanostructures present at the surface. The inset in (c) shows the RHEED pattern of the surface of the sample taken at room temperature after growth. (e) Morphology of a 70-nm-thick Bi$_2$Te$_3$ film grown on a (0001) Al$_2$O$_3$ substrate at 230\textdegree C and of a (f) 20-nm-thick Bi$_2$Te$_3$ film grown on an STO substrate at 250\textdegree C. From the line profile obtained across a series of wide steps (blue line), the step height was determined to be 1 QL (right-hand panels) for samples grown on both substrates. Reproduced and adapted from~\cite{PhD_Tthesis_Ngabonziza}.}
 \label{fig:Fig-02}
\end{figure*}

To passivate the substrate surface and achieve atomically sharp interfaces between the TI film and the substrate, the two-step growth scheme for  VdW epitaxy of Bi$_2$Te$_3$ is often used~\cite{SEHarrison_2013,PNgabonziza_2015,LHe_2013,VMPereira_2021}. After initial treatments of the substrate, a few nanometers of  Bi$_2$Te$_3$ are first deposited at a slightly lower substrate temperature. This initial low-temperature layer, referred to as the nucleation layer, effectively fulfills the objective of passivating the substrate's dangling bonds for the subsequent high-temperature VdW epitaxy. The lower-temperature layer is then annealed as the substrate temperature is raised to higher growth temperatures, and the second layer is epitaxially deposited on top. This two-step growth procedure often results in Bi$_2$Te$_3$ thin films of high-quality, terrace-step surface morphology with fewer defects, and also with larger and fewer triangular mounds, see Fig.~\ref{fig:Fig-02}. Furthermore, adopting the two-step growth method has been demonstrated to facilitate the growth of bulk insulating Bi$_2$Te$_3$ films on a wide variety of substrates~\cite{PNgabonziza_2015,KHoefer_2014,TPGinley_2016,VMPereira_2021}.

For practical purposes, large terraces are preferred, the size of which strongly depends on the substrate temperature during growth, which must be chosen properly. To establish the correct substrate temperature ($T_{\text{sub}}$) for MBE growth of the Bi$_2$Te$_3$ films, we have performed a set of depositions at different $T_{\text{sub}}$ for samples grown on different substrates. Representative samples grown on both sapphire and (111)-oriented  (STO) substrates are presented in Fig.~\ref{fig:Fig-02}. Using the two-step growth method, we found that the optimal growth temperature for Bi$_2$Te$_3$ prepared on sapphire is  170\textdegree C for the nucleation layer and 230\textdegree C for the subsequent layer. For samples grown on STO with the two-step method as well, the optimal temperature for the nucleation layer is also 170\textdegree C, whereas for the second layer it is 250\textdegree C. Figures~\ref{fig:Fig-02}\textcolor{blue}{a}--\textcolor{blue}{d} demonstrates the improvement of the surface morphology as $T_{\text{sub}}$ is increased for four different samples of 15-nm Bi$_2$Te$_3$ grown on STO [111]. 

In practice, if $T_{\text{sub}}$ is too low, an amorphous or a polycrystalline film will be formed, as observed in the film grown at 170\textdegree C, see Fig.~\ref{fig:Fig-02}\textcolor{blue}{(a)}. This is because the adatoms will not have enough energy to diffuse and find the lowest potential energy site. As the substrate temperature is slightly increased from 170 to 210\textdegree C, an improved morphology with the appearance of triangular mounds is observed, see Fig.~\ref{fig:Fig-02}\textcolor{blue}{(b)}. However, at this temperature, there are still notable small and large  3D or 1D defects, indicated by arrows in Fig.~\ref{fig:Fig-02}\textcolor{blue}{(b)}~\cite{SEHarrison_2013}. 
The same features were also reported for MBE growth of Bi$_2$Se$_3$ films, a prototypical material of Bi$_2$Te$_3$ ~\cite{ZYWang_2011,TPGinley_2016}. For the sample grown at 250\textdegree C and annealed for 5 min in Te flux (after growth of the second layer), the 3D and 1D defects disappear and the surface is dominated by triangular mounds as depicted in Fig.~\ref{fig:Fig-02}\textcolor{blue}{(c)}. 
The corresponding reflection high-energy electron diffraction (RHEED) image, see inset in Fig.~\ref{fig:Fig-02}\textcolor{blue}{(c)}, taken after cooling to room temperature, signals that the film is highly oriented. Annealing samples longer ($\sim$30 min) helps produce a smooth surface,  Fig.~\ref{fig:Fig-02}\textcolor{blue}{(d)}. However, it should be noted that longer annealing times often cause Te to re-evaporate, resulting in Te-deficient films or non-stoichiometric samples, depending on the sticking coefficient of Te on different substrates at high temperatures~\cite{AMzerd_1995,AMzerd_1994}.

The morphology of the resulting films at optimal substrate temperatures demonstrated characteristic triangular-shaped terraces and steps, see Figs.~\ref{fig:Fig-02}\textcolor{blue}{(e)}--\textcolor{blue}{(f)}. The surface morphology of pure Bi$_2$Te$_3$ films typically exhibits this type of spiral-like triangular domains~\cite{JKrumrain_2011,SMorelhao_2017,CIFornari_2016,HSteiner_2014,SEHarrison_2013}. Such triangular islands highlight the fcc stacking for the rhombohedral crystal structure of Bi$_2$Te$_3$~\cite{ZYWang_2011,LHe_2013}. The lateral size of these triangular mounds is $\sim$1.5~µm for films grown  on both sapphire and STO. The adjacent terraces were separated by an average step height of $\sim$1~nm, consistent with the thickness of a single QL in the [001] direction of Bi$_2$Te$_3$, see right-hand panels, extracted line profiles in Figs.~\ref{fig:Fig-02}\textcolor{blue}{(e)} and \textcolor{blue}{(f)}~\cite{HZhang_2009}. From a surface morphology point of view on a small scale, namely the lateral size of triangular mounds and the step height of terraces, the difference in growth on the two substrates is insignificant, in agreement with previous reports~\cite{LHe_2013,CIFornari_2016,HSteiner_2014,SEHarrison_2013}.

The formation of spiral-like structures is observed at the top of the triangular island, see Figs.~\ref{fig:Fig-02}\textcolor{blue}{(e)}and \textcolor{blue}{(f)}. Previous studies on binary (Bi$_2$Te$_3$ and Bi$_2$Se$_3$) and (off-) stoichiometric ternary Bi-based TI thin films have shown the formation of spirals for samples grown on a variety of substrates~\cite{HDLi_2010,LHe_2013,TPGinley_2016,SEHarrison_2013}. At first glance, the spirals' growth behavior in Bi-based TI systems might be 
considered to come from misfit dislocations at the interface during epitaxial growth, as in conventional semiconductor systems~\cite{AKarma_1998,YCui_2002,XDong_2016}, where the spirals are often considered to come from misfit dislocations at the interface during heteroepitaxial growth. When the dislocations have a screw component, they can emerge at the surface to form additional steps to facilitate the 
growth of spirals.  However, spiral growth without dislocations in Bi-based TI systems has been demonstrated~\cite{YLiu_2012,GHao_2013}, where the adjacent QLs are only weakly bonded by VdW forces, and the epilayer is not expected to bond strongly to any substrate, i.e., no formation of dislocations is anticipated to seed the spiral growth. Furthermore, Wang \textit{et al.}~\cite{KWang_2013} have reported the absence of spirals on the terraces of Bi$_2$Te$_3$ films. Thus, contrary to dislocation-mediated spiral growth, the observed spirals in Bi$_2$Te$_3$ thin films most likely form as a result of the pinning of the 2D growth fronts at the substrate steps or domain boundaries during the initial and intermediate stages of growth, respectively~\cite{YLiu_2012}. 

\begin{flushleft} 
\textit{2.2 Correlation between Surface Morphology and Structural Properties}
\end{flushleft}
\paragraph*{} 
Different studies have investigated the correlation between surface morphology and structural defects of epitaxial Bi$_2$Te$_3$ films grown on different substrates,  which demonstrated that such defects depend strongly on the substrate material and growth conditions~\cite{CIFornari_2018,DKriegner_2017,JKampmeier_2015,CI_Fornari_2020}. 

\begin{figure*}[!t]
\centering
\includegraphics[width=1\textwidth]{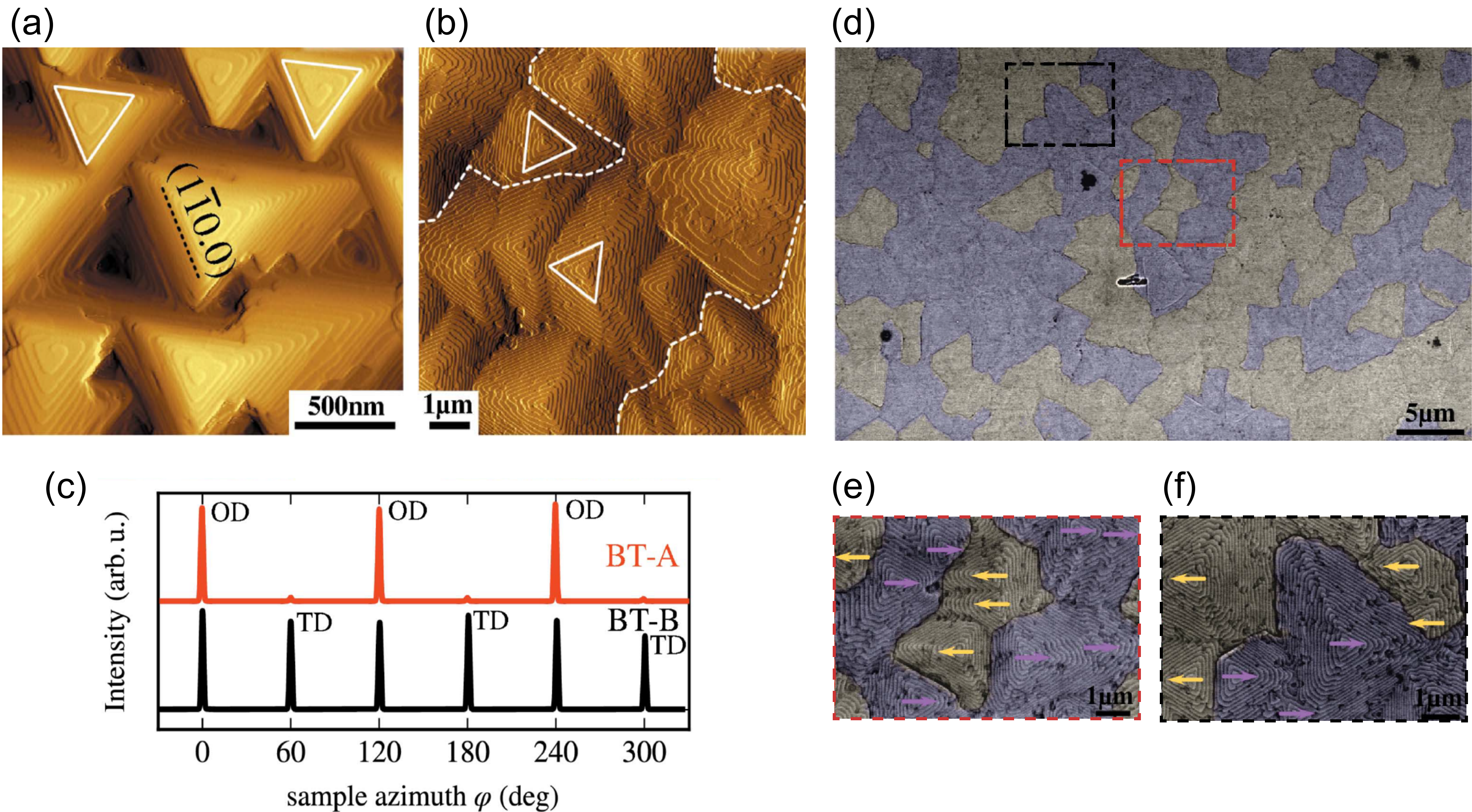}
\caption{\linespread{1.5} Surface morphology and structural properties correlations for 200-nm-thick Bi$_2$Te$_3$ films grown on (111)-oriented BaF$_2$ substrates. Surface morphology of two Bi$_2$Te$_3$ samples grown at different substrate temperatures: (a) $T_{\text{sub}}=280$\textdegree C and (b) $T_{\text{sub}}=320$\textdegree C. The lattice plane indices of the preferred surface step orientation are indicated in (a). Both samples show characteristic triangular-shaped pyramids on the surface, but the sample grown at $T_{\text{sub}}=320$\textdegree C demonstrates pyramids with two orientations rotated by 60/180\textdegree. Dashed lines in (b) mark the boundary between areas with the two distinct orientations. The side steps of the pyramids are oriented along $(1 \bar{1}0.0)$ lattice planes as indicated in (a).
(c) X-ray diffraction azimuth scans highlighting  a threefold symmetry according to the crystal structure for the sample grown at $T_{\text{sub}}=280$\textdegree C (red), whereas the sample grown at $T_{\text{sub}}=320$\textdegree C (black) shows a sixfold symmetry. The origin of the Bragg peaks from the ordinary domains (OD) and twin domains (TD) is indicated. (d) SEM image of the sample in (b) overlaid with a semi-transparent representation of the electron backscatter diffraction crystal orientation data. In agreement with XRD data shown in (c), only two types of orientations, i.e. the two twin orientations, are observed. The TD boundaries are correlated with the surface structure as shown in enlarged SEM images (e) and (f). Arrows indicate the orientation of the triangular-shaped pyramids. All areas shown in yellow show the triangular pyramids pointing to the left, whereas the areas shown in violet point to the right. Reprinted from~\cite{DKriegner_2017}, \copyright ~2017 International Union of Crystallography.}
 \label{fig:Fig-03}
\end{figure*}

By performing a combined bulk- and surface-sensitive study of the epitaxial Bi$_2$Te$_3$ films grown on (111)-oriented BaF$_2$, Kriegner \textit{et al.}~\cite{DKriegner_2017} resolved buried TDs not visible on the surface of the samples. Depending on the growth temperature of Bi$_2$Te$_3$ films, the orientation of the triangle mounds varied. Triangular pyramidal features with edges in two different orientations rotated in the surface plane by 60\textdegree ~were observed, see Figs.~\ref{fig:Fig-03}\textcolor{blue}{(a)} and \textcolor{blue}{(b)}.  The dissimilarities were attributed to the different growth temperatures, which also resulted in clear differences to the structural properties of samples grown at different $T_{\text{sub}}$, see Fig.~\ref{fig:Fig-03}\textcolor{blue}{(c)}. For the sample grown at $T_{\text{sub}}=280$\textdegree C, a predominantly threefold symmetry was observed, highlighting an almost complete absence of any twinning, in agreement with the surface morphology with only one orientation of the surface pyramids. Whereas for samples grown at $T_{\text{sub}}=320$\textdegree C, a strong twinning with almost equal abundance of the ordinary domain (OD) and twin domain (TD) regions was observed in the film. This observation was also in  agreement with combined scanning electron microscopy (SEM) and electron backscatter diffraction crystal orientation data, see Figs.~\ref{fig:Fig-03}\textcolor{blue}{(d)} and \textcolor{blue}{(f)}, which further confirmed that the surface topography is indeed correlated with the local crystal orientation of the thin film. 

\begin{figure*}[!t]
\centering
\includegraphics[width=1\textwidth]{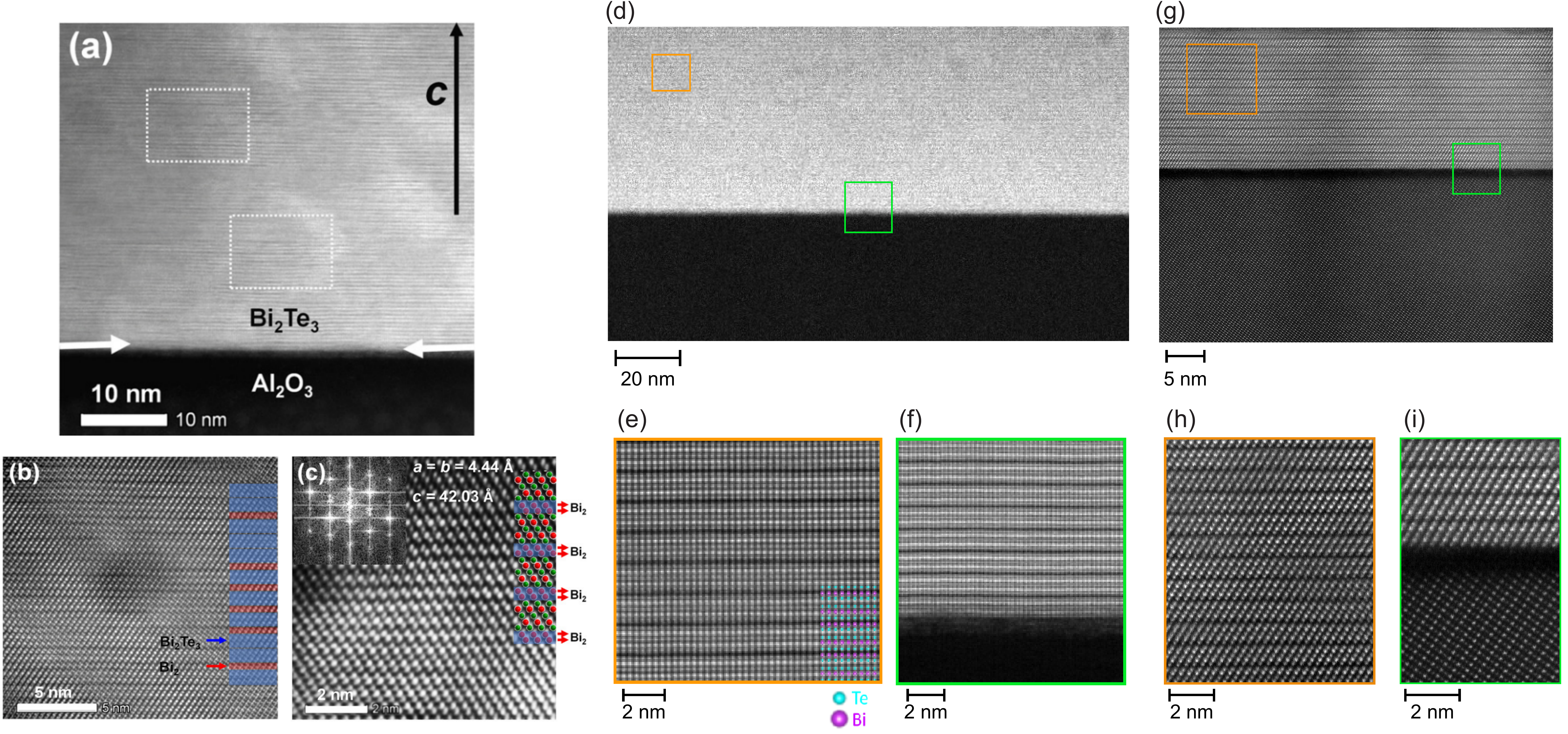}
\caption{\linespread{1.5} Scanning transmission electron microscope  (STEM) characterizations of Bi$_2$Te$_3$ films grown on different substrates under distinct growth conditions. (a)--(c) STEM images showing the cross-sectional structures for  Bi$_2$Te$_3$ films grown at $T_{\text{sub}}= 270$\textdegree C on an Al$_2$O$_3$ $(00l)$ substrate: (a) Low-magnification TEM image of a Bi$_2$Te$_3$ film, the white dotted boxes indicate the local distorted regions; (b) and (c) show the atomic-resolution STEM images in the marked areas. The inset in (c) shows the corresponding fast Fourier transformation image from the sequential stack of Bi$_2$-Bi$_2$Te$_3$. Overview STEM images of Bi$_2$Te$_3$ films grown using a two-step temperature growth on (d) Al$_3$O$_3$ [0001] and (g) (111)-oriented STO substrates. Magnified STEM images of the samples showing: (e,h) QLs of the films and (f,i) a smooth interface between the films and the substrates. The bottom right-hand inset in (e) displays a schematic atomic structural model showing the sequence of Te and Bi atoms in a quintuple layer of Bi$_2$Te$_3$. (a)--(c) Reprinted from \cite{MZhang_2021} with  permission of AIP Publishing. (d)--(f) Reprinted and partly adapted with permission from \cite{PNgabonziza_2018} \copyright ~2018,  American Physical Society.}
 \label{fig:Fig-04}
\end{figure*}

On the other hand, it is also worth mentioning some beneficial effects of the formation of special defects in epitaxial Bi$_2$Te$_3$ films. As recently indicated for Bi-rich  Bi$_2$Te$_3$ films grown on sapphire substrates, it was observed that, when the Bi excess is sufficiently high,  $n$-type Bi$_2$ planar defects form in the VdW gap, which were clearly identified in the cross-sectional structural characteristics of such samples, see Figs.~\ref{fig:Fig-04}\textcolor{blue}{(a)}--\textcolor{blue}{(c)}. As these Bi$_2$ defects are excellent electron donors, enhancement of the electron density by over one order of magnitude was observed, accompanied by the suppression of carrier-intrinsic excitations. It was also established that such Bi-rich Bi$_2$Te$_3$ films with Bi$_2$ planar defects possess a much improved TE power compared to that of the similar Bi$_2$Te$_3$ films without Bi$_2$ intercalations~\cite{MZhang_2021}.

By using an extremely low growth rate of 2.7~nm~h$^{-1}$, Kampmeier \textit{et al.}~\cite{JKampmeier_2015} managed to suppress previously reported TDs in epitaxial Bi$_2$Te$_3$ films grown on (111)-oriented Si substrates~\cite{SBorisova_2013}. The deposited layer followed the threefold symmetry of the substrate surface, and high-quality single crystalline and TD-free Bi$_2$Te$_3$ films were prepared~\cite{JKampmeier_2015}. This finding of achieving TD-free films corroborates  our microstructural analysis using scanning transmission electron microscopy (STEM) for Bi$_2$Te$_3$ films grown on sapphire~\cite{PNgabonziza_2018} and STO substrates. No noticeable domain boundaries, propagating through the complete films, were resolved, see Figs.~\ref{fig:Fig-04}\textcolor{blue}{(d)}--\textcolor{blue}{(i)}. For both samples, no misfit dislocations generating edge-type defects or threading dislocations extending vertically through the film were observed, see Figs.~\ref{fig:Fig-04}\textcolor{blue}{(d)} and \textcolor{blue}{(g)}. Instead, highly parallel QLs are clearly visible, which are separated by VdW gaps, see Figs.~\ref{fig:Fig-04}\textcolor{blue}{(e)} and \textcolor{blue}{(h)}.  A closer view of the interfacial region between the film and the substrates reveals highly parallel layers  in the film close to the interface, despite the large lattice mismatch between Bi$_2$Te$_3$ [001] and these substrates ($\sim$8.7\% for sapphire and $ -10.8\%$ for STO [111])~\cite{LHe_2013}. This suggests a high crystal quality at the film--substrate interface, see Figs.~\ref{fig:Fig-04}\textcolor{blue}{(f)} and \textcolor{blue}{(i)}. Moreover, contrary to a previous report on
Bi$_2$Te$_3$ films grown on Si substrates~\cite{SBorisova_2013}, no dislocations and distortions of the atomic column were observed immediately above the substrate, confirming a rapid relaxation of strain at the interface. 

Finally, it is also necessary to note that, in the practice of preparing stoichiometric Bi$_2$Te$_3$ films with better transport characteristics, it is important to ascertain that the correct phase has been grown. There are reports on Bi$_2$Te$_3$ films grown on BaF$_2$ substrates indicating that using small Te/Bi beam equivalent pressure ratios during growth led to the formation of bismuth bilayers between QLs~\cite{CIFornari_2016,OCaha_2013,AFulop_2014}. This resulted in the formation of either  BiTe or Bi$_4$Te$_3$ phases in addition to or in place of the desired Bi$_2$Te$_3$ phase, and neither of these phases show evidence of the TI surface state. Furthermore, samples with such extra phases have considerably lower electron mobilities and higher carrier concentrations than pure Bi$_2$Te$_3$ films. Thus, the ability to control the Bi$_2$Te$_3$ film growth is  essential for obtaining high-quality films with  low defect density and  high electron mobility in order to explore their transport properties 
for the electrons in the surface states, as discussed in the next section.

\section*{\Large{3. Q\lowercase{uantum} T\lowercase{ransport in} B\lowercase{i$_2$}T\lowercase{e$_3$} t\lowercase{hin} f\lowercase{ilms} }}
\paragraph*{}
As most of the fascinating applications of TI materials are for electronic devices,  a fundamental understanding of their electronic transport properties is needed before they can be implemented in novel electronic devices. For electronic transport characterizations of most 3D TIs, especially for Bi$_2$Te$_3$ samples, bulk conduction complicates the direct exploitation of the surface effect. A very promising approach is to improve the crystal quality by carefully controlling the growth conditions of the films. This allows us to obtain samples with fewer intrinsic defects and high surface mobilities~\cite{YLi_2010,PNgabonziza_2015,KHoefer_2014,KHoefer_2015}. Moreover, thin films 
allow  the sample thickness to be controlled accurately in order to investigate the thickness-dependent transport properties, thus achieving the thickness range for observing a crossover from a 3D TI to a 2D TI, in which the films are in the quantum-spin Hall (QSH) transport regime~\cite{CXLiu_2010,YZhang_2010}. Furthermore, (i) the possibility of preparing \textit{in situ} films and capping them immediately after growth to avoid extrinsic defects, (ii) the ability of interfacing films with other materials to make TI heterostructures, and (iii) the flexibility of preparing several electronic devices and multilayers on the same film render thin films much more advantageous than bulk single crystals for quantum transport studies. 
MBE-grown thin films are also especially advantageous compared to exfoliated films because they offer the ability to dope the materials in a controllable manner, which led, for example, to the observations of the quantum anomalous Hall effect in magnetic TIs~\cite{CZChang_2013,CZChang_2015,CZChang_2013-02}.

From a quantum-transport perspective, studying transport properties in an externally applied magnetic field is a powerful tool for   investigating the topological nature of the surface state. From magnetotransport experiments, transport characteristics such as the carrier density and mobility in the film can be extracted. Moreover, interesting phenomena such as nonlinear ordinary Hall effect \cite{PNgabonziza_2018,ZZhang_2012},  planar Hall effect~\cite{ABhardwaj_2021}, weak antilocalization \cite{ARoy_2013,HTHe_2011} and Shubnikov--de Haas (SdH) oscillations~\cite{DPHolgado_2020,XYu_2012} have been reported in magnetotransport experiments on Bi$_2$Te$_3$ thin films. In addition,  micro- and nano-structured Bi$_2$Te$_3$ thin films demonstrated gate-tunable changes in transport properties, the full depletion of bulk carriers, as well as the successful realization of combined gate-tunable Josephson supercurrents and normal-state transport of Josephson junctions and Hall bar devices fabricated side-by-side~\cite{PNgabonziza_2016,MPStehno_2020}.

\begin{figure*}[!t]
\centering
\includegraphics[width=1\textwidth]{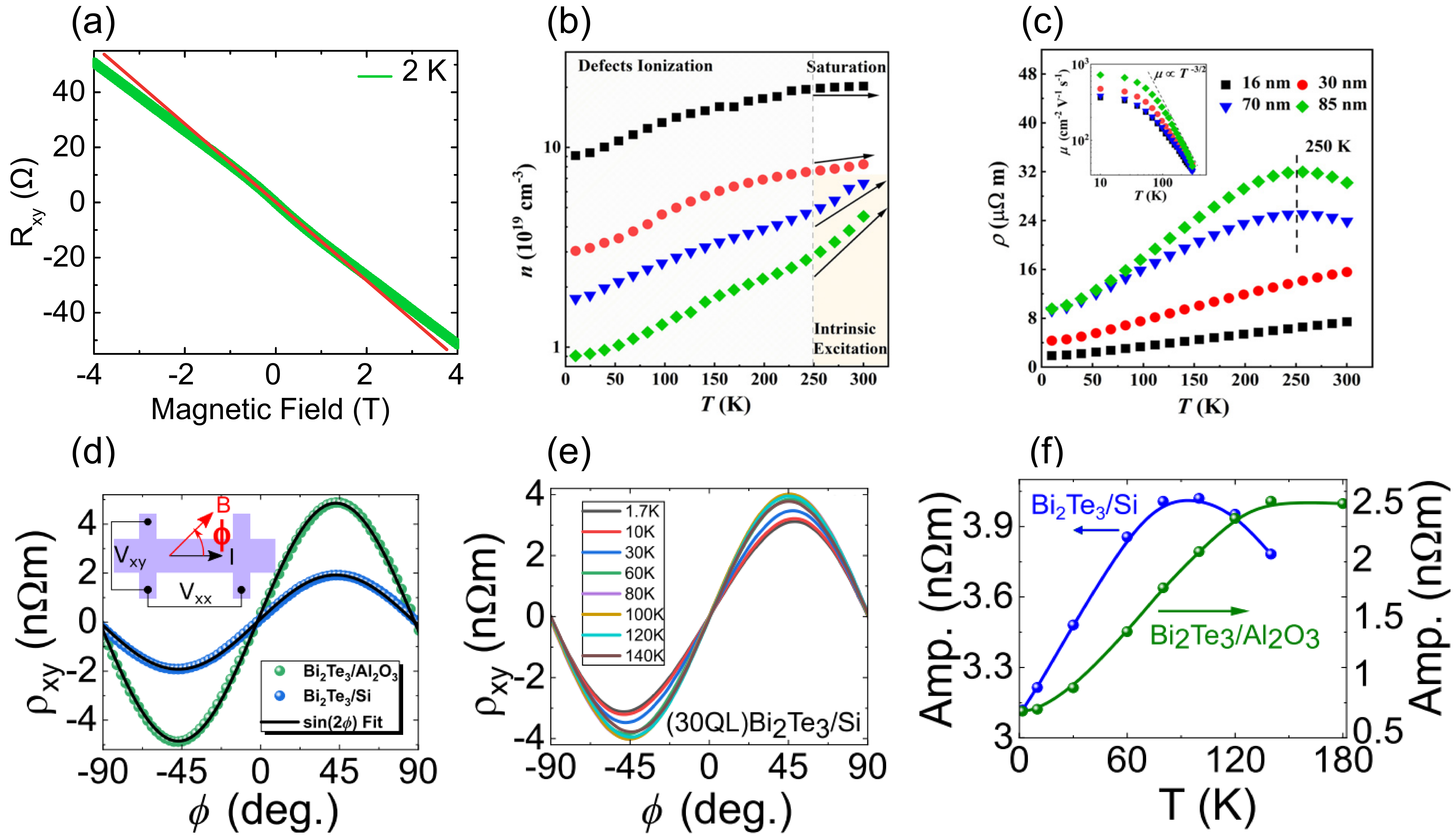}
\caption{\linespread{1.5} Ordinary and planar Hall effects in Bi$_2$Te$_3$ thin films. (a) Nonlinear ordinary Hall effect in a 70-nm Bi$_2$Te$_3$ film grown on sapphire $(000l)$ substrates using a two-step temperature growth: The Hall resistance versus magnetic field at 2~K plotted together with a one-carrier model fitting to highlight the presence of multiple bands carrier transport. (b) Temperature-dependent  electron concentration extracted from the ordinary Hall signal and (c) electrical resistivity for Bi$_2$Te$_3$ films grown at 245\textdegree C with different thicknesses. The inset in (c) depicts the temperature-dependent carrier mobility of the films. (d) Planar Hall transverse resistivity ($\rho_{xy}$) versus azimuthal angle ($\phi$) for 14-nm-thick films of Bi$_2$Te$_3$ grown on Si and sapphire substrates plotted together with a $\sin 2\phi$ fit. The inset depicts the schematic illustration of the measurement geometry. (e) $\rho_{xy}$ versus $\phi$ at different temperatures of a  30-nm-thick film of Bi$_2$Te$_3$ grown on Si. (f) The PHE amplitude versus temperature for 30-nm-thick films of Bi$_2$Te$_3$ grown on Si and sapphire substrates. (a)  Reprinted and partly adapted with permission from \cite{PNgabonziza_2018}, \copyright ~2018 American Physical Society. (b)--(c) Reprinted from \cite{MZhang_2020}; and (d)--(f) from \cite{ABhardwaj_2021} with  permission of AIP Publishing.}
 \label{fig:Fig-05}
\end{figure*}

\begin{flushleft} 
\textit{3.1 Ordinary Hall Effect and Planar Hall Effect}
\end{flushleft}
\paragraph*{} 
The ordinary Hall effect is the development of a transverse electric field in a solid material when it carries an electric current. This requires a magnetic field to be perpendicular to both the electric current direction and the sample plane. Experimentally, the nonlinear ordinary Hall effect is commonly measured in transport experiments on 3D TI films at low temperatures~\cite{HSteinberg_2011,AATaskin_2012,PNgabonziza_2018,ZRen_2010}. In particular, the nonlinearity in the ordinary Hall signal $R_{xy}(B)$ has been measured in Bi$_2$Te$_3$ TI films, Bi$_2$Te$_3$-based heterostructures,  thin films of its prototypical material Bi$_2$Se$_3$, as well as in hybrid Bi$_2$Se$_3$/Bi$_2$Te$_3$ heterostructures~\cite{PNgabonziza_2018,NBansal_2012,ZZhang_2012,YZhao_2013}. This nonlinearity in the $R_{xy}(B)$ signal has been suggested to originate from the coexistence of bulk and surface transport channels. If all carriers participating in transport had the same mobility, $R_{xy}(B)$ would show linear behavior with the slope determined by $1/(eR_\textrm{H})$, where $R_\textrm{H}$ is the Hall coefficient. However, when there are multiple types of carriers with different but comparable electronic mobilities, nonlinearity is observed in the $R_{xy}(B)$ signal, see Fig.~\ref{fig:Fig-05}\textcolor{blue}{(a)}. For electronic characterizations of TIs, these ordinary Hall measurements also provide an opportunity to get a much clearer picture of the carrier density and mobility in the films than those determined, for example, from ARPES analysis. Table~\ref{table_1} provides a representative overview of reported electrical properties from transport measurements on Bi$_2$Te$_3$ films grown on different substrates~\cite{AFulop_2014,PNgabonziza_2016,PNgabonziza_2018,JKampmeier_2015,ARoy_2013,MPStehno_2020,KWang_2013,MZhang_2020,YZhao_2013}. 

To determine an accurate  contribution of the surface states to the overall electronic transport properties, it is crucial to understand the behavior of the bulk of the sample. This was emphasized recently in a detailed analysis of the carrier densities and electronic mobilities extracted from the $R_{xy}(B)$ signal of low-defect-density Bi$_2$Te$_3$ films: High-mobility bulk electrons were identified to contribute significantly to the overall magnetotransport properties~\cite{PNgabonziza_2018}. Ideally, thinner samples minimize the contribution of bulk modes in electronic transport by decreasing the total number of dopant charges, thus indicating that the surface transport through the topological surface states will play an increasingly predominant role when the film thickness is decreased. However, structural defects such as intrinsic point defects could have strong negative effects on the overall measured transport properties of the film. M.~Zhang~\textit{et al.}~\cite{MZhang_2020} observed a monotonous decrease of $n_{3\textrm{D}}$ extracted from the ordinary Hall signal when  increasing the thickness of Bi$_2$Te$_3$ films, leading to an obvious increase in resistivity and mobility, see Figs.~\ref{fig:Fig-05}\textcolor{blue}{(b)} and \textcolor{blue}{(c)}. These behaviors were attributed  to \textit{in situ} transformation of point defects, which highlights the particular role played by point defects for modulating and controlling the electronic transport properties of Bi$_2$Te$_3$ films. 

Next, the planar Hall effect (PHE) in 3D TI materials is discussed. In the past few years, the PHE in topological materials has evoked extensive interest. Contrary to the ordinary Hall effect, the PHE manifests itself as a detectable transverse voltage in response to a longitudinal current, under an external in-plane applied magnetic field, in a configuration where the transverse voltage due to the Lorentz force is zero. Interest in this effect in topological materials has been stimulated by its role in probing topological characteristics (PHE has been proposed to originate from surface-state conduction). Taskin~\textit{et al.}~\cite{AATaskin_2017} demonstrated that the PHE amplitude measured in Bi$_{2-x}$Sb$_x$Te$_3$ thin films  across the Fermi level  exhibited local maxima on either side of the Dirac point in the surface-state regime. Somewhat similar observations were also reported in bulk crystals of Sn-doped Bi$_{1.1}$Sb$_{0.9}$Te$_2$S, which showed oscillating PHE characteristics only in the topologically protected surface-state regime~\cite{BWu_2018}. Although  both the linear~\cite{BWu_2018,AATaskin_2017} and nonlinear~\cite{PHe_2019,KYasuda_2017} PHE in TIs have been confirmed experimentally to originate from surface states,
little is known about their microscopic mechanisms. Various 
microscopic mechanisms have been proposed to  explain this phenomenon theoretically~\cite{WRao_2021,SHZheng_2020}.

\begin{table}[!t]
\caption{Electrical transport properties at low temperatures ($<$50~K) for Bi$_2$Te$_3$ thin films grown on different substrates. The electron mobility $(\mu)$ as well as the sheet density $(n_\textrm{2D})$ and bulk density $(n_\textrm{3D})$ are shown wherever data were available.}
\centering
\begin{tabular}{|C{ 8 cm}|C{2 cm}|C{2 cm}|C{ 2.5 cm}|C{ 1cm}|}
\hline
 Details of material and substrates & $\mu$\newline [cm$^2$/Vs]& $n_\textrm{2D}$\newline [cm$^{-2}$]  & $n_\textrm{3D}$\newline [cm$^{-3}$] & Refs. \\ \hline
 &&&&\\
Bi$_2$Te$_3$ on Si $[111]$ & 1030 & --& $4.7 \times 10^{19}$& \cite{AFulop_2014} \\ 
 &&&&\\
 Bi$_2$Te$_3$ (capped with Al$_2$O$_3$) on SrTiO$_3 \, [111]$ &1594 &$3.8 \times 10^{12}$ & -- & \cite{PNgabonziza_2016}\\
  &&&&\\
  Bi$_2$Te$_3$ (capped with Al$_2$O$_3$) on Al$_2$O$_3 \, [0001]$ &1206 &$9.5 \times 10^{12}$ & -- & \cite{PNgabonziza_2016}\\
   &&&&\\
   Bi$_2$Te$_3$ on Al$_2$O$_3 \, [0001]$ &2800 &$5.8 \times 10^{13}$ & $8.2 \times 10^{18}$ & \cite{PNgabonziza_2018}\\
    &&&&\\
Bi$_2$Te$_3$ on pre-structured Si $[111]$  &110 &$1.6 \times 10^{14}$ & -- & \cite{JKampmeier_2015}\\
 &&&&\\
Bi$_2$Te$_3$ on $n$-type Si $[111]$ &35 &$1.2 \times 10^{14}$ & -- & \cite{ARoy_2013}\\
 &&&&\\
Bi$_2$Te$_3$ on SrTiO$_3 \, [111]$  &700 to 1900 &$2.8 \times 10^{13}$ & -- & \cite{MPStehno_2020}\\
 &&&&\\
Bi$_2$Te$_3$ on mica  &5800 &$2.45 \times 10^{12}$ & -- & \cite{KWang_2013}\\
 &&&&\\
Bi$_2$Te$_3$ on Al$_2$O$_3 \, [0001]$  & $\sim$1000 &-- & $0.5 \text{ to } 10\times 10^{19}$ & \cite{MZhang_2020}\\
 &&&&\\
Bi$_2$Te$_3$ on Al$_2$O$_3 \, [0001]$  & $912$ &-- & $ 4.17\times 10^{19}$ & \cite{YZhao_2013}\\
 &&&&\\
  \hline

\end{tabular}

\label{table_1}
\end{table} 

A.~Bhardwaj \textit{et al.}~\cite{ABhardwaj_2021} recently reported  linear  PHE in Bi$_2$Te$_3$ thin films grown on Si and sapphire substrates. As expected for the linear PHE, the planar Hall transverse resistivity $\rho_{xy}$ varied as $\sin 2\phi$ with a period of $\pi$, see Fig.~\ref{fig:Fig-05}\textcolor{blue}{(d)}. Contrary to the reported experimental linear PHE in Bi$_{2-x}$Sb$_x$Te$_3$ thin films~\cite{AATaskin_2017}, the PHE amplitude was observed to increase with increasing temperature for thicker (30~nm) films, see Figs.~\ref{fig:Fig-05}\textcolor{blue}{(e)} and \textcolor{blue}{(f)}. Unlike the extensive observation of the ordinary Hall effect in 3D TI thin films, only a few publications have very recently reported the PHE in nonmagnetic Bi-based 3D TI thin films~\cite{AATaskin_2017,PHe_2019,ABhardwaj_2021}. In particular, the linear PHE has not yet been reported experimentally in Bi$_2$Se$_3$; only nonlinear unconventional PHE with a periodicity of $2\pi$ has been reported in Bi$_2$Se$_3$ films~\cite{PHe_2019}. This further emphasizes the need for in-depth  experimental exploration and theoretical understanding of the PHE in thin films and heterostructures of Bi-based 3D TIs. For a detailed comprehensive discussion of the linear and nonlinear PHE in TIs, see~\cite{WRao_2021,SHZheng_2020,SNandy_2018}.

\begin{flushleft} 
\textit{3.2 Shubnikov--de Haas Oscillations}
\end{flushleft}
\paragraph*{} 
Quantum oscillations are determined by Landau quantization of the energy levels of crystalline solids in magnetic fields due to the density of states being periodically modulated as a function of magnetic field~\cite{shoenberg_1984}. In particular, the oscillations occurring in conductivity are called Shubnikov--de Haas (SdH) oscillations. Conductivity oscillates periodically as a function of the inverse of the magnetic field $(1/B)$. The oscillatory part of the longitudinal conductivity is given by
$\Delta \sigma_{xx}\sim \cos \left[ 2\pi \left( \frac{F}{B}-\frac{1}{2}+\phi\right] \right)$, where $F$ is the oscillation frequency and $\phi$ is the phase factor $(0< \phi <1)$. The phase factor $\phi$ of the SdH oscillations directly reflects the Berry phase, thus providing an opportunity to check whether the electrons leading to these oscillations are Dirac fermions.  Experimentally, $\phi$ can be determined by carefully analyzing the so-called Landau level (LL) fan diagram. For Dirac systems with a linear energy dispersion $(\phi = 1/2)$, a topologically non-zero Berry phase of $\pi$ is expected~\cite{AATaskin_2011-1,GPMikitik_2012}, whereas for spinless fermions with a parabolic energy dispersion $(\phi = 0)$, the Berry phase is zero~\cite{shoenberg_1984,GPMikitik_1999}.  Therefore, SdH oscillations have been used for the quantitative characterization and careful disentanglement of  the 2D TI surface states that coexist with 3D bulk states, particularly  also because their angular dependence in tilted magnetic fields provides information about the size, shape and dimensionality of the Fermi surface (FS)~\cite{ELahoud_2013,LBao_2012}.

SdH oscillations have been used as a powerful probe of the nature of the metallic state in Bi$_2$Te$_3$ TI samples~\cite{MVeldhorst_2012,MVeldhorst_2013,KWang_2013}. Notably, the electron mobility $\mu$ of Bi$_2$Te$_3$ thin films has been reported to be sufficiently high enough to fulfill the condition $\mu B \gg 1$ for allowing SdH oscillations in the resistance at high magnetic fields to be detected~\cite{PNgabonziza_2018,DPHolgado_2020,KWang_2013,DRosenbach_2020,HJin_2015}. 
For Bi$_2$Te$_3$ films, reported SdH oscillations have been interpreted to originate from surface states, from   high-mobility bulk electrons, or from additional topologically trivial states due to Rashba splitting in the transport characteristics of Bi$_2$Te$_3$ FS~\cite{PNgabonziza_2018,KWang_2013,DPHolgado_2020}. 

\begin{figure*}[!t]
\centering
\includegraphics[width=1\textwidth]{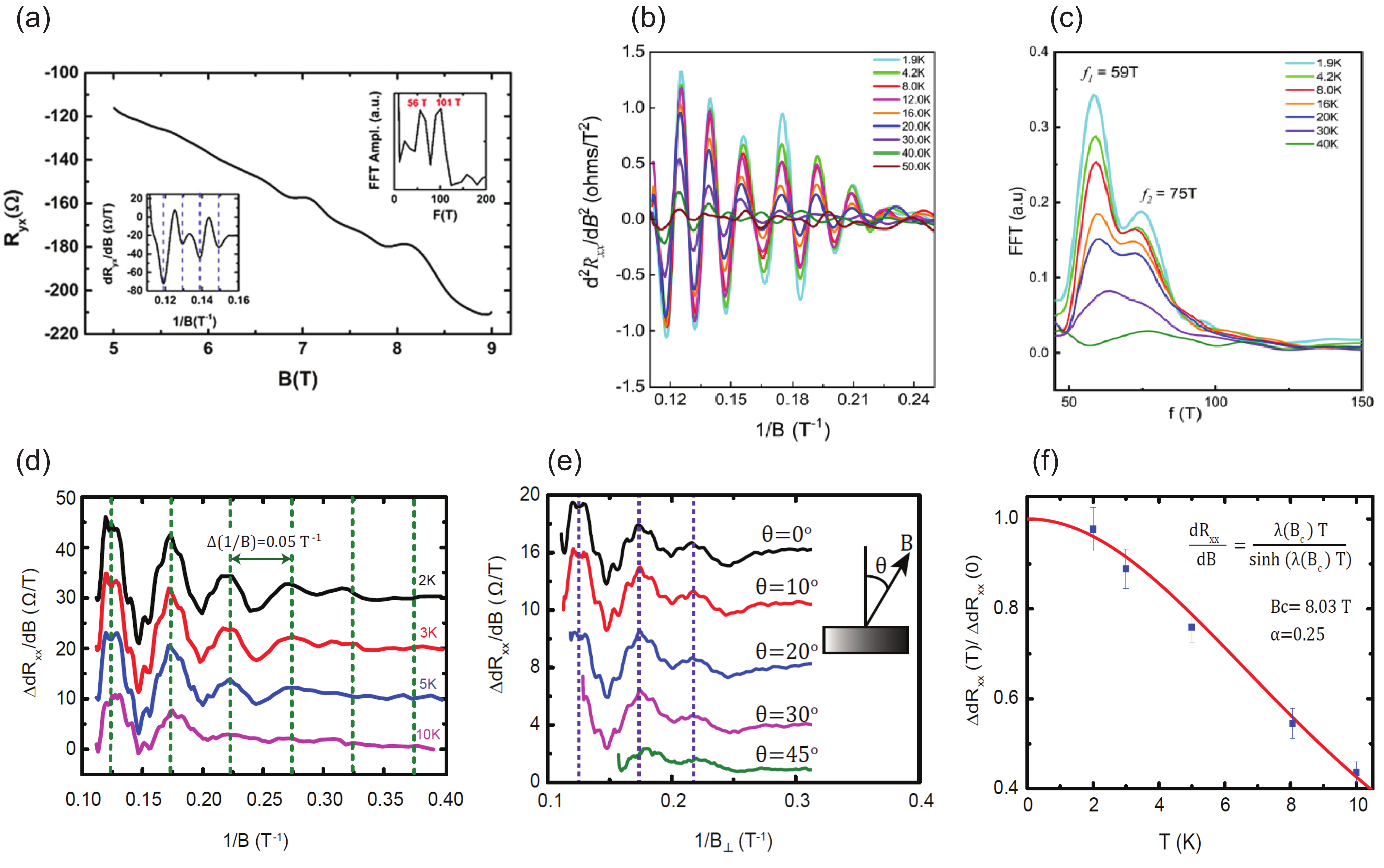}
\caption{\linespread{1.5} Shubnikov--de Haas (SdH) oscillations in Bi$_2$Te$_3$ films grown on different substrates. (a) SdH oscillations from a 4-nm-thick film grown on a mica substrate: The Hall resistance $R_{yx}$ at 2~K shows quantum oscillations. The lower and upper insets depict the first derivative of $R_{yx}$ against $1/B$ and the fast Fourier transform (FFT) of $dR_{yx}/B$, respectively. (b)~SdH oscillations for a 156-nm-thick film grown on a BaF$_2$ (111) substrate. The oscillations were extracted from the second derivative of $R_{xx}$ with respect to the applied magnetic field and (c)~FFT curves from the SdH oscillations at various temperatures. SdH oscillations in a 70-nm-thick film grown on a sapphire $(000l)$ substrate: (d)~The oscillatory part of the longitudinal resistance $\Delta R_{xx}/dB$ as a function of $1/B$. (e)~SdH oscillations for different tilt angles $\theta$ plotted as a function of $1/B_{\perp}$ with $B_{\perp}= B\cos (\theta)$. The inset shows the direction of the applied magnetic field. (f)~Temperature dependence of the normalized longitudinal resistance, where the solid red line is the best fit to the function $\lambda (T)/\sinh (\lambda (T))$.  (a)~Reprinted from~\cite{KWang_2013}; (b)--(c)~reprinted from \cite{DPHolgado_2020} with permission of AIP Publishing. (d)--(f)~Reprinted from \cite{PNgabonziza_2018}, \copyright ~2018 American Physical Society.}
 \label{fig:Fig-06}
\end{figure*}

Initial quantum transport experiments on Bi$_2$Te$_3$ single crystals and thin films detected SdH oscillations, which were often attributed to originate from topological surface states based on a cosine magnetic field-angle dependence of the oscillations and the magnetic field dependence of the LLs~\cite{MVeldhorst_2012,KWang_2013,DXQu_2010,SBarua_2014}.  For example, in an early analysis of SdH oscillations for Bi$_2$Te$_3$ thin films, K.~Wang \textit{et al.}~\cite{KWang_2013} reported two-frequency SdH oscillations, which they attributed to originate from the top and bottom surface states with different carrier densities, see Fig.~\ref{fig:Fig-06}\textcolor{blue}{(a)}. From the LLs analysis, however, deviations from the expected $\pi$-Berry phase were observed, which were attributed to arise from the interference between the two oscillation frequencies. Nevertheless, such a non-zero Berry phase could also originate from a non-topological trivial band that crosses the Fermi energy due to band bending at the surface, as observed for Bi$_{1.5}$Sb$_{0.5}$Te$_{1.7}$Se$_{1.3}$ samples~\cite{AATaskin_2011}. More specifically, this could also be due to Rashba splitting states that break the spin degeneracy and give rise to the observed  two frequencies~\cite{DPHolgado_2020}.

Indeed, Holgado \textit{et al.}~\cite{DPHolgado_2020} recently reported  SdH oscillations in Bi$_2$Te$_3$ epitaxial films, and their systematic study demonstrated that such SdH oscillations did not originate from the topological surface states, but rather emerge from the Rashba splitting of the FS of Bi$_2$Te$_3$. Figure~\ref{fig:Fig-06}\textcolor{blue}{(b)} depicts SdH oscillations from the second derivative of longitudinal resistance as a function of $1/B$. Two well-developed peaks with frequencies of $f_1=59$~T and $f_2=75$~T were extracted from the corresponding fast Fourier transform (FFT) analysis, see Fig.~\ref{fig:Fig-06}\textcolor{blue}{(c)}.
There are several possible interpretations of these two frequencies. The first is to consider that the electronic transport through the Bi$_2$Te$_3$ film is composed of normal fermions from the bulk of the sample and of Dirac fermions from topological surface states. The second is that these two frequencies originate from the top and bottom surface states, as concluded from early quantum transport experiments on Bi-based 3D TI samples~\cite{MVeldhorst_2012,KWang_2013,MPetrushevsky_2012,TRDevidas_2014,CZhang_2014}. However, the angular dependence of the measured oscillations in a tilted magnetic field on the same sample did not follow the expected 2D FS state behavior. Thus, it was concluded that the measured SdH oscillations  originated from 3D bulk states. In particular, it was indicated in this study that the Rashba effect is responsible for the two frequencies observed in the FFT, a proposal that is in agreement with previous reports on several systems (e.g., bulk BiTeCl, SnTe epilayer and HgTe quantum wells)~\cite{FXXiang_2015,AKOkazaki_2018,CRBecker_2003}. The origin of such Rashba splitting in Bi$_2$Te$_3$ epitaxial films was attributed to the intrinsic disorder present in the sample or due to surface oxidation of the Bi$_2$Te$_3$ films. The latter is supported by the fact that Bi$_2$Te$_3$ samples are known to experience surface chemical modifications upon exposure to ambient conditions~\cite{PNgabonziza_2018,HBando_2000}. There is also theoretical support indicating that adsorbed atoms from the atmosphere alter the surface states of Bi$_2$Te$_3$ by inducing Rashba splitting due to the charge transfer from the adsorbed atoms to the sample surface~\cite{KHJin_2012}.

Regarding the 2D nature of the SdH oscillations, which is often taken as evidence of surface-state transport in 3D TIs, it should be emphasized in the case of Bi$_2$Te$_3$ that SdH oscillations should be analyzed meticulously. Owing to the cylindrical shape of the Bi$_2$Te$_3$ bulk FS, the angle dependence of the bulk magnetoresistance oscillations is two-dimensional in nature, thus highlighting  the possibility of detecting 2D quantum oscillations from high-mobility bulk electrons from such a cylindrical FS. Indeed, our recent magnetotransport study on low-defect-density Bi$_2$Te$_3$ films showed that SdH oscillations arise from the high-mobility bulk electrons, see Fig.~\ref{fig:Fig-06}\textcolor{blue}{(d)}. However, the maxima and minima of the  measured SdH oscillations were aligned with each other for different angles when oscillations are plotted as a function of the perpendicular component of the magnetic field ($B_{\perp}=B \cos(\theta)$), see Fig.~\ref{fig:Fig-06}\textcolor{blue}{(e)}~\cite{PNgabonziza_2018}. A careful comparison of the Hall carrier density with the spherical and cylindrical FS carrier densities from SdH oscillations demonstrated that the FS has a shape between that of a spherical FS and an open cylindrical FS, but more elongated towards a cylindrical FS. This indicates that the measured quantum SdH oscillations originate from bulk carriers. Figure~\ref{fig:Fig-06}\textcolor{blue}{(f)} depicts the temperature dependence of the normalized longitudinal resistance. By analyzing  SdH oscillations at different temperatures using the Lifshitz--Kosevich expression and the Dingle plot~\cite{shoenberg_1984,YAndo_2013,TDietl_1978}, a surface mobility was extracted from SdH oscillations of $\mu_s=3600$~cm$^2$ V$^{-1}$s$^{-1}$, which is consistent with the mobility value extracted from the Hall data. This agreement between bulk and surface mobilities, as determined from Hall and SdH data, respectively, further indicates that bulk states significantly contribute to the measured SdH oscillations in magnetotransport~\cite{PNgabonziza_2018}. As discussed above for SdH oscillations originating from the Rashba splitting effect, which are due most probably to surface oxidation~\cite{DPHolgado_2020}, and because our Bi$_2$Te$_3$ films were exposed to air for days during device fabrication processes, it is reasonable to assume that there are chemical surface modification effects that effectively dope electrons into the bulk. This implies that electronic conductivity of the film is no longer determined by the surface states alone.  These observations highlight that, for further \textit{ex situ} electronic device fabrication steps and subsequent quantum transport studies, Bi$_2$Te$_3$ films should first be capped \textit{in situ} immediately after growth, before exposure to air and other \textit{ex situ} contaminations. This will protect the surface states from degradation and unintentional doping.

\begin{flushleft} 
\textit{3.3 Protective Capping and Electrostatic Tuning of the Fermi Level}
\end{flushleft}
\paragraph*{} Usually, device-structuring processes and quantum-transport measurements are not performed in vacuum systems connected to the UHV chambers used for combined \textit{in situ} TI thin-film deposition and characterizations. Therefore, it is important to develop methods for \textit{in situ} capping of TI films to avoid any possible contamination and protect the TI surface states prior to \textit{ex situ} device fabrication steps and subsequent quantum transport experiments. An advanced multicluster UHV system architecture with multiple deposition and analysis chambers for \textit{in situ} synthesis, capping and surface characterization  is ideal for detailed investigations before and after \textit{in situ} capping of the films. Preferably, such a multicluster UHV system  should also be connectable to a vacuum suitcase for long-distance transport of samples in UHV conditions for further \textit{in situ} analysis to other facilities such as synchrotron radiation facilities. Hoefer \textit{et al.}~\cite{KHoefer_2015,VMPereira_2021} and Ngabonziza \textit{et al.}~\cite{PNgabonziza_2016,PNgabonziza_situ_2018} have independently employed variants of such multicluster UHV systems for film growth, \textit{in situ} capping and characterizations of high-quality, bulk-insulating Bi$_2$Te$_3$ epitaxial thin films. 

\begin{figure*}[t]
\centering
\includegraphics[width=1\textwidth]{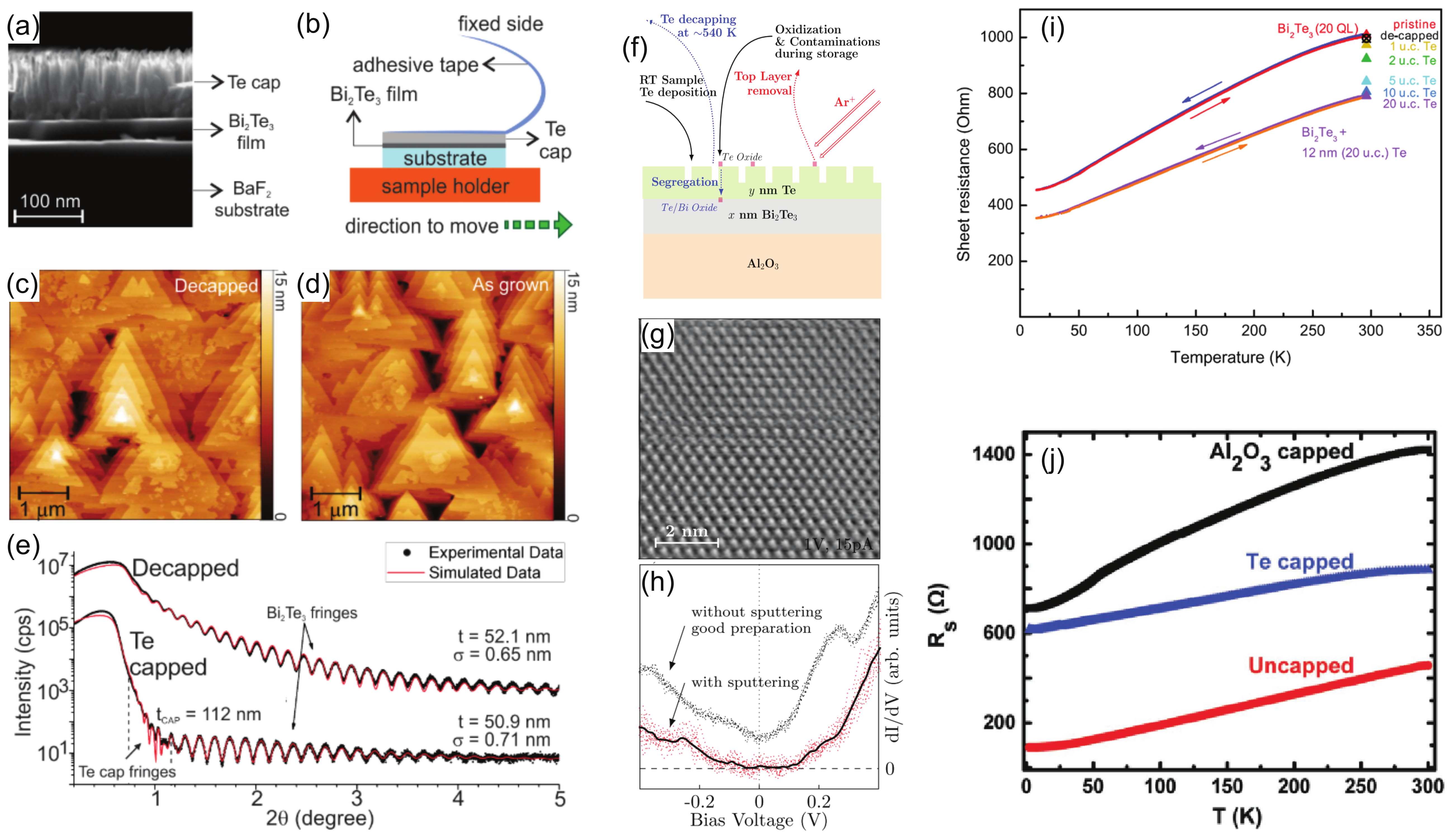}
\caption{\linespread{1.5} Protective capping and decapping of epitaxial Bi$_2$Te$_3$ films. (a)~Cross section field-emission gun microscopy image of a 
25-nm-thick Bi$_2$Te$_3$ film grown on $(111)$-oriented BaF$_2$ substrates and capped with a 100-nm Te layer. (b)~Illustration of the mechanical capping layer removal technique that uses adhesive tape. Atomic force microscopy images of (c)~decapped and (d)~as-grown 25-nm-thick Bi$_2$Te$_3$ films. (e)~X-ray reflectivity of a Bi$_2$Te$_3$ film before and after decapping procedure. Recovering the pristine surface of 30-nm-thick Bi$_2$Te$_3$ films grown on sapphire substrates after long-time storage: (f)~Illustrative image of the Te capping and decapping process, highlighting the need for doing  argon-ion etching before decapping. (g)~STM image showing the atomically resolved structure of the tellurium-terminated surface layer. The pristine surface is recovered after argon-ion sputtering of the long-time stored Bi$_2$Te$_3$ film. (h)~Scanning tunneling spectroscopy spectra from the same surface (solid line) and after decapping (dots). Electronic transport characteristics of \textit{in situ}-capped Bi$_2$Te$_3$ films: (i)~\textit{In situ} sheet resistance measurements of the pristine and Te-capped 20-nm-thick Bi$_2$Te$_3$ films grown on $(111)$-oriented BaF$_2$ substrates; (j)~Temperature dependence of the sheet resistance for \textit{ex situ}-structured Hall bar devices on three different 15-nm-thick  Bi$_2$Te$_3$ films grown on $(111)$-oriented STO substrates (black: capped with Al$_2$O$_3$, blue: capped with Te and red: uncapped film). (a)--(e)~Reprinted from~\cite{CIFornari_2016}. (f)--(h)~Reprinted from~\cite{LULiang_2021}. (i)~Reprinted from from~\cite{KHoefer_2015}. (j)~Reprinted with permission from~\cite{PNgabonziza_2016}, \copyright ~2016 John Wiley and Sons. }
 \label{fig:Fig-07}
\end{figure*}

For \textit{in situ} capping of Bi$_2$Te$_3$ films, the first approach has been to identify a suitable capping layer that would preserve the surface states and avoid unintentional doping induced by ambient conditions. Systematic characterizations  of the surface properties of intrinsically insulating Bi$_2$Te$_3$ films before and after \textit{in situ} capping with an epitaxially grown Te layer  have been reported~\cite{PNgabonziza_2016,KHoefer_2015}. This epitaxially grown Te layer could be an appropriate capping material because the surface of Bi$_2$Te$_3$ is naturally Te-terminated. Moreover, it is a removable capping layer within the temperature range of Bi$_2$Te$_3$ growth.  An amorphous layer of Te~\cite{SEHarrison_2014,CIFornari_2016},  Se~\cite{KVirwani_2014}, or an evaporated layer of Al~\cite{XYu_2012} have also been used as a protective capping layer for Bi$_2$Te$_3$ films. However, there have been questions related to whether the intrinsic topological properties of Bi$_2$Te$_3$ films are influenced by such an amorphous capping overlayer. For example, alterations of the stoichiometry after  removal of a Se/Te capping layer have been reported~\cite{SEHarrison_2014,KVirwani_2014}. In addition, an Al$_2$O$_3$ capping layer grown by atomic layer deposition was reported to potentially cause damage to the Bi$_2$Te$_3$ surface states~\cite{HLiu_2011}. 

Nevertheless, the successful removal of Te capping layers on Bi$_2$Te$_3$ films has been reported~\cite{CIFornari_2016,LULiang_2021,KHoefer_2015}. Fornari~\textit{et al. }~\cite{CIFornari_2016} demonstrated the possibility to deposit a protective Te capping layer immediately after  Bi$_2$Te$_3$ epitaxial growth and a mechanical decapping process to expose fresh surfaces by using the adhesive tape technique, as commonly used for cleaving bulk TI single crystals, see Figs.~\ref{fig:Fig-07}\textcolor{blue}{(a)} and \textcolor{blue}{(b)}. Systematic surface and structural characterizations after applying this Te decapping strategy to Bi$_2$Te$_3$ epitaxial films demonstrated that pristine surface and crystal structure are totally preserved, see Figs.~\ref{fig:Fig-07}\textcolor{blue}{(c)}--\textcolor{blue}{(e)}. 
Recently, Liang~\textit{et al.}~\cite{LULiang_2021} reported the successful use of a protective Te capping layer for long-time storage and transfer of an epitaxial Bi$_2$Te$_3$ film as well as a temperature-controlled method for removing the Te capping layer, which involved argon-ion etching of the film surface  prior to annealing for decapping, see Fig.~\ref{fig:Fig-07}\textcolor{blue}{(f)}. After argon-ion sputtering, the Bi$_2$Te$_3$ pristine surface is recovered at a decapping temperature of $\sim$550~K,  Figs.~\ref{fig:Fig-07}\textcolor{blue}{(g)} and \textcolor{blue}{(h)}, suggesting that nearly perfect TI surfaces can be reliably attained even after long-time storage through a combination of an initial argon-ion sputtering process, followed by heating for decapping. 

Hoefer \textit{et al.}~\cite{KHoefer_2015} and Ngabonziza \textit{et al.}~\cite{PNgabonziza_2016} conducted systematic studies of the influence of a Te capping layer on the electrical transport properties of intrinsically bulk-insulating Bi$_2$Te$_3$ films. Comparative \textit{in situ} investigations of the sheet resistance were reported for pristine Bi$_2$Te$_3$ films and for the film fully capped with a Te layer. After the Te capping step, the sheet resistance decreased gradually, indicating the contribution of the Te capping layer to the measured electronic conductivity, see Fig.~\ref{fig:Fig-07}\textcolor{blue}{(i)}~\cite{KHoefer_2015}. Temperature-dependent sheet resistance measurements on \textit{in situ} Te-capped Bi$_2$Te$_3$ films, which were  structured \textit{ex situ} into Hall bar devices, also indicated a small contribution of the thin Te protective cap to the transport properties of the Bi$_2$Te$_3$~\cite{PNgabonziza_2016}. Therefore, for systematic and in-depth quantum transport investigations, an insulating capping layer would be preferable as it would help avoid a possible contribution of conduction from the capping material in measured TI electronic transport characteristics. For example, the Al$_2$O$_3$-capped Bi$_2$Te$_3$ film was reported to show the highest sheet resistance compared to uncapped and \textit{in situ} Te-capped Bi$_2$Te$_3$ films, Fig.~\ref{fig:Fig-07}\textcolor{blue}{(j)}. This suggests that the insulating  Al$_2$O$_3$ capping layer would be more appropriate for the direct exploitation of the surface states in quantum transport experiments~\cite{PNgabonziza_2016}. Furthermore, such an insulating protective capping layer offers the opportunity to achieve a dual-gating configuration in thin-film devices fabricated from \textit{in situ}-capped Bi$_2$Te$_3$ samples for efficient electrostatic tuning of the Fermi level.

\begin{figure*}[t]
\centering
\includegraphics[width=1\textwidth]{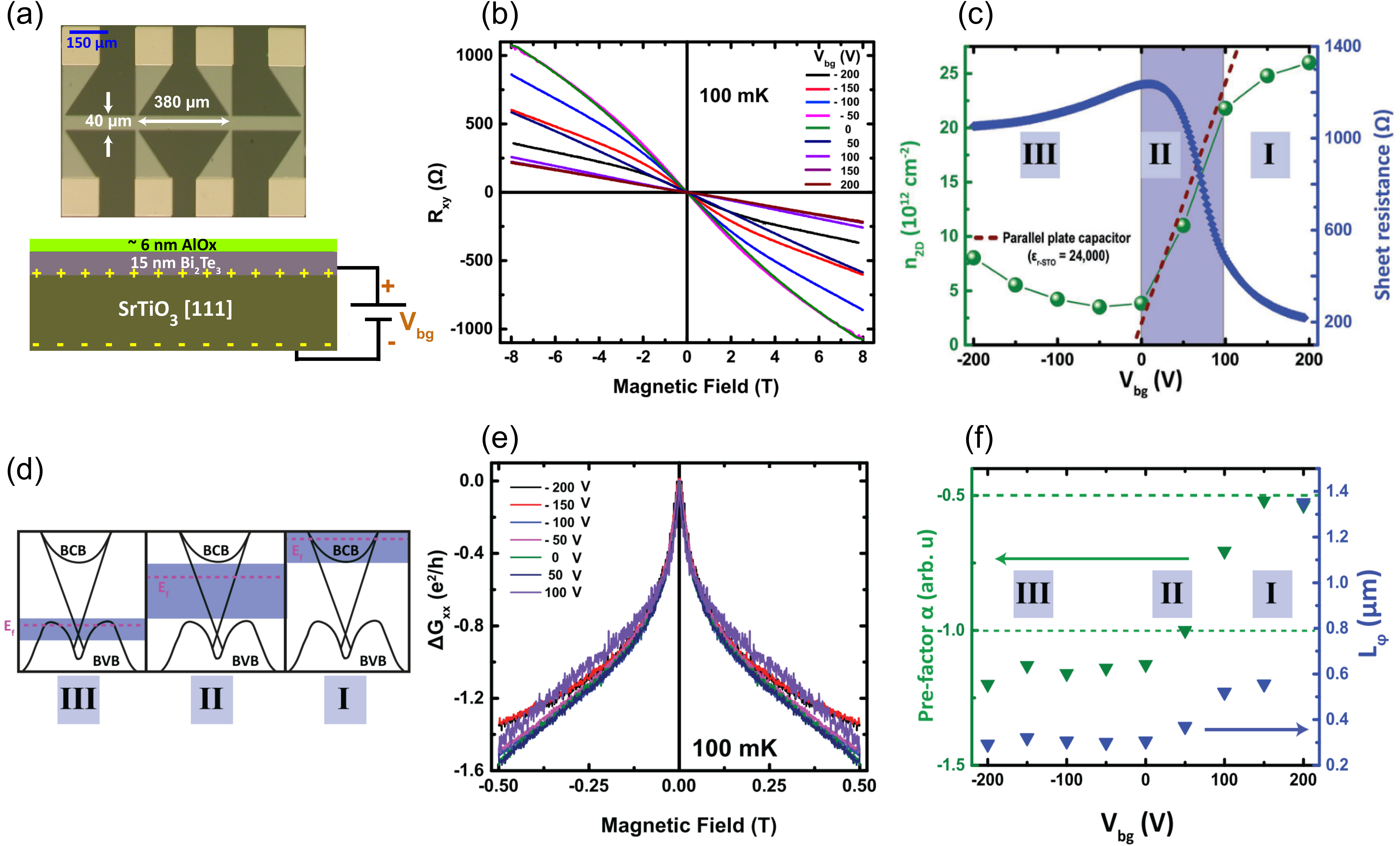}
\caption{\linespread{1.5} Electronic transport characteristics of \textit{in situ}-capped 15-nm-thick Bi$_2$Te$_3$ films  grown on a (111)-oriented STO  substrate. (a)~Top panel: Optical micrograph image of the structured Hall bar device. Bottom panel: Schematic representation of a typical \textit{in situ}-capped Bi$_2$Te$_3$ sample, where the total thickness of the A$_2$O$_{\text{x}}$ capping layer is $\sim$6~nm.  (b)~Back-gate voltage dependence of the Hall resistance, $R_{xy}$ at 100~mK. (c)~2D carrier density extracted from the Hall resistance zero-field slope of $R_{xy}$ in (b) and the sheet resistance as a function of the 
applied back-gate voltage $V_{\text{bg}}$. (d)~Schematic band diagram illustration in which BCB is the bulk conduction band, BVB is the bulk valence band, and $E_{\text{F}}$ is the position of the Fermi level. The transport regimes I, II, and III are accessed  in (c) as the Fermi level moves through the respective shaded regions of the band structure. (e)~Back-gate dependence of the weak antilocalization effect, $\Delta G_{xx}(B)$, for \textit{in situ} Al$_2$O$_3$-capped Bi$_2$Te$_3$ sample.  (f)~Extracted pre-factor $\alpha$ and dephasing length $\l_{\phi}$ at different back-gate voltages. Reprinted and partly adapted with permission from~\cite{PNgabonziza_2016},  \copyright ~2016 John Wiley and Sons. }
 \label{fig:Fig-08}
\end{figure*}

We have demonstrated the effectiveness of combining \textit{in situ} capping using an insulating protective cap layer and electrostatic tuning of the Fermi level in Bi$_2$Te$_3$ thin-film devices~\cite{PNgabonziza_2016}. This combined capping and electrostatic gating strategy provides a technological platform to investigate the topological properties of intrinsically insulating Bi$_2$Te$_3$ surface states in quantum transport. This paves the way towards realizing a variety of Bi$_2$Te$_3$-based topological quantum devices. Figure~\ref{fig:Fig-08}\textcolor{blue}{(a)} depicts a representative structured Bi$_2$Te$_3$ Hall bar device and a schematic illustration of a typical \textit{in situ}-capped Bi$_2$Te$_3$ film grown on a (111)-oriented STO substrate, where the substrate is used as the back-gate dielectric. Magnetotransport properties of   Al$_2$O$_3$-capped Bi$_2$Te$_3$ films were modulated by applying back-gate voltages in the range of $\pm 200$~V. For example, the slope of the Hall resistance  ($R_{xy}$) at different back-gate voltages ($V_{\text{bg}}$) was observed to increase with decreasing $V_{\text{bg}}$. Distinct nonlinearity in $R_{xy} (B)$ was observed for $V_{\text{bg}}\leq 50$~V, see Fig.~\ref{fig:Fig-08}\textcolor{blue}{(b)}. Qualitatively similar gate-tunable magnetotransport characteristics were reported in different Bi-based TI films grown on (111)-oriented STO substrates~\cite{XHe_2013,GZhang_2011,JChen_2011,JChen_2010}. In the analysis of the response of sheet carrier density and sheet resistance with respect to $V_{\text{bg}}$, three different gate voltage regions were identified with distinct characteristics, see Fig.~\ref{fig:Fig-08}\textcolor{blue}{(c)}. At high positive gate voltages (region I), the Fermi level is located in the bulk conduction band. Observations of a maximum in the sheet resistance at $V_{\text{bg}}\sim 10$~V, which is followed by a downturn and saturation at negative gate voltages, were interpreted as an increase in interband scattering when the Fermi level touches the bulk valence band~\cite{PNgabonziza_2016}. This is because the Dirac point of Bi$_2$Te$_3$ is buried below the bulk valence band maximum. Thus the resistance maximum is not correlated with the depletion of the surface states at the Dirac point. In the gate voltage range of $\pm 200$~V, the 2D carrier density was modulated by a factor of $\sim$7, and the observed low carrier concentration in the intermediate region was considered to be a strong indication that the Fermi level is located in the bulk band gap, crossing only the topologically nontrivial metallic surface states, see Fig.~\ref{fig:Fig-08}\textcolor{blue}{(d)}. At high negative gate voltages, the carrier density starts to increase slowly again as the bulk valence band contributes mobile carriers. These results demonstrate the possibility of achieving full depletion of bulk carriers in the Al$_2$O$_3$-capped Bi$_2$Te$_3$ films, thus allowing access to the topological transport regime dominated by surface-state conduction.

Weak antilocalization (WAL) measurements at different gate voltages further support the observed carrier distribution in Al$_2$O$_3$-capped Bi$_2$Te$_3$ films~\cite{PNgabonziza_2016}. WAL is a quantum effect that is commonly observed in TI systems due to the spin-momentum locking resulting from strong spin-orbit coupling~\cite{HZLu_2014}. The $\pi$-Berry phase of the topological surface states leads to destructive interference between time-reversed charge carrier paths, which results in an increase of measured conductivity at zero magnetic field. The quantum correction to the 2D conductivity due to the WAL effect was calculated by Hikami, Larkin, and Nagaoka (HLN)~\cite{SHikami_1980}: 
$\Delta G(B)= \alpha\frac{e^2}{\pi h}\left[ \Psi \left( \frac{B_{\varphi}}{B}+\frac{1}{2} \right) -\ln \left(\frac{B_{\varphi}}{B} \right)\right].$
The pre-factor $\alpha$ accounts for the WAL contributions from different conducting channels. From the  HLN equation, the pre-factor $\alpha$ is expected to be $-1/2$ for a single 2D conducting surface with strong spin-orbit coupling. If there are two independent 2D conducting channels, the pre-factors add up. Contributions from either a conducting bulk or disorder-induced electron--electron interactions in the system will cause the value of $\alpha$ to deviate from $-1/2$ towards more positive values~\cite{SPChiu_2013,JWang_2011}. An effective dephasing length $l_{\varphi}$ will replace the phase relaxation length of the individual channel~\cite{JLee_2012,IGarate_2012}. For elaborated theoretical and experimental discussions of the WAL effect in TI systems, the reader is referred to~\cite{MBrahlek_2015,HLu_2011,IGarate_2012,JChen_2010,SPChiu_2013}.

In capped Bi$_2$Te$_3$ and other Bi-based TI thin films, the WAL effect has been investigated at different gate voltages~\cite{PNgabonziza_2016,FYang_2014,YHLiu_2015}. Figure~\ref{fig:Fig-08}\textcolor{blue}{(e)} depicts representative sheet conductance $\Delta G_{xx}(B)$ as a function of magnetic field at various $V_{\text{bg}}$ for Al$_2$O$_3$-capped Bi$_2$Te$_3$ films~\cite{PNgabonziza_2016}. These $\Delta G_{xx}$ data show WAL behavior. The extracted pre-factor $\alpha$ and dephasing length $l_{\varphi}$ show modulations for a wide range of applied $V_{\text{bg}}$, see Fig.~\ref{fig:Fig-08}\textcolor{blue}{(f)}. 
At high positive $V_{\text{bg}}$ (region I), the value of $\alpha$ is $\sim -0.5$, and it drops to $\sim -1$ in the intermediate regime (region II). It then saturates at a value slightly lower than $- 1$ for negative back-gate voltages (region III). The same trend was observed for the $V_{\text{bg}}$ dependence of the dephasing length. These WAL data on Al$_2$O$_3$-capped Bi$_2$Te$_3$  thin-film devices further demonstrate the possibility of using a back-gate to tune the Fermi level all the way from the bulk conduction band to the topological surface states, and then to the bulk valence band. The ability to prepare Bi$_2$Te$_3$ thin films  with low intrinsic doping, paired with protective capping of the topological surface  and effective depletion of bulk carriers by back gating, is an important step toward implementing a variety of topological Bi$_2$Te$_3$-based quantum  devices. 

\begin{flushleft} 
\textit{3.4 Proximity-Induced Superconductivity in Bi$_2$Te$_3$ Thin-Film Devices}

\end{flushleft}
\paragraph*{}
Next, recent progress in quantum transport studies on hybrid structures, consisting of Bi$_2$Te$_3$ thin films interfaced to an $s$-wave superconductor in topological Josephson junction devices, is discussed. The field of topological superconductivity has drawn considerable interest in the condensed-matter research community. It offers the unique prospect of combining properties of  band-structure topology with superconducting pairing  to generate exotic topological states, for example Majorana bound states (MBS)~\cite{LFu_2008,YTanaka_2009,ACPotter_2011}. MBS in condensed-matter systems could potentially be used as topological qubits to perform fault-tolerant computations because they obey non-abelian statistics~\cite{CNayak_2008,AKitaev_2003}. Recently, signatures of MBS have been experimentally observed in
various materials~\cite{VMourik_2012,SMAlbrecht_2016,ADFinck_2013,JWiedenmann_2016,EBocquillon_2017,Nadj-Perge_2014}. Common to all these experiments is that superconducting correlations are induced by bringing the material in contact with an ordinary $s$-wave superconductor. For an elaborated introduction to topological superconductivity and Majorana fermions in condensed-matter physics, the interested reader can consult available references on this topic~\cite{CBeenakker_2013,MLeijnse_2012,MSato_2017,MSato_2016}.

\begin{figure*}[t]
\centering
\includegraphics[width=1\textwidth]{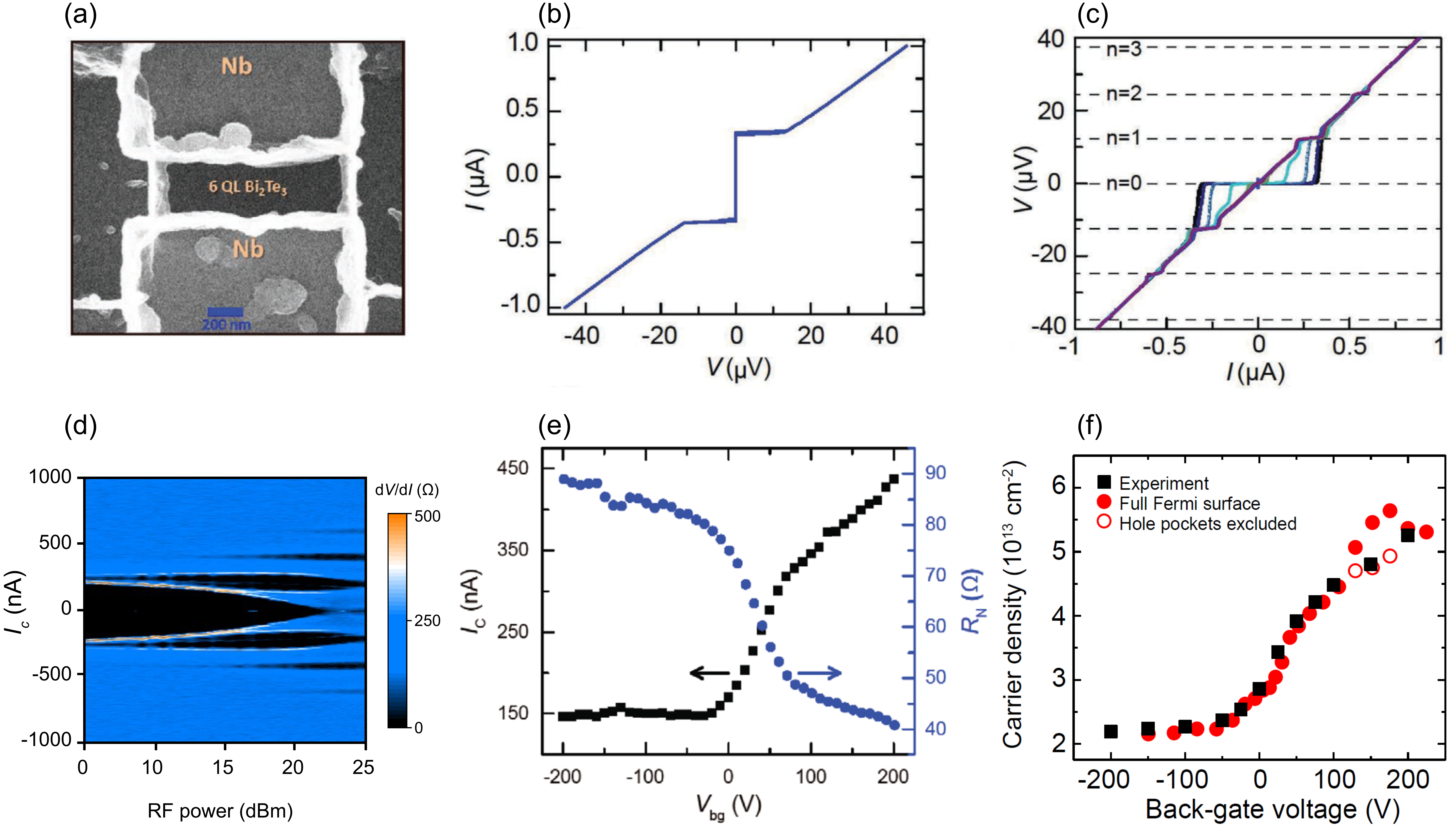}
\caption{\linespread{1.5} Electronic transport characteristics in Nb$/$Bi$_2$Te$_3/$Nb Josephson junction devices. (a)~SEM image of a typical Josephson junction device. The device dimensions are clearly visible by residues of cross-linked photoresist from the etch mask. (b)~The \textit{I--V} characteristic of a Josephson device with a 6-nm-thick weak link, and (c)~the measured voltage plateaus appear in the \textit{V--I} curves under microwave irradiation with a frequency of 6.02~GHz, as well as (d)~the corresponding color-scale plot of the differential resistance $(dV/dI)$ as a function of the DC and rf current drives  $I$ and $I_\textrm{rf}$, respectively (at zero back-gate voltage). The black region in the color-scale maps corresponds to the voltage plateaus in the \textit{V--I} curves, see panel (c). (e)~Back-gate voltage dependence of the normal state resistance and the Josephson critical current in the same Josephson device with a 6-nm-thick weak link. (f)~Experimental carrier density and modeling of the carrier distribution in a 15-nm-thick Bi$_2$Te$_3$ film: total carrier density as a function of gate voltage from Hall measurements (black squares) and from band structure modeling (red filled circles). (a)--(c) and (e)--(f)~Reprinted from~\cite{MPStehno_2020}.}
 \label{fig:Fig-09}
\end{figure*}

A promising approach in the search for MBS is to use 3D TI-based Josephson junctions. Early on, theoretical proposals were put forward  to use a network of Josephson junctions on the surface states of a 3D TI in order to realize an artificial topological superconductor that resembles a spinless chiral $p_x + i p_y$ superconductor without breaking time-reversal symmetry~\cite{LFu_2008}. Subsequent theoretical work demonstrated that this proximity-induced superconductivity has an orbital term of the $p+s$-wave type, in which the $p$-wave part is composed of conventional chirals $p_x + i p_y$ and $p_x - i p_y$ that add to the $s$-wave term to form the total superconducting order parameter~\cite{GTkachov_2013}.
The first observations of Josephson supercurrent in 3D TI were reported in devices fabricated from Bi-based single crystals~\cite{BSacepe_2011,MVeldhorst_2012}. In these single crystals, most studies demonstrated in the Josephson transport that a large fraction of the supercurrent is carried by ballistic Andreev bound states, which was attributed to the topological surface state~\cite{LGalletti_2014,MVeldhorst_2012}. However, no other signatures were observed, and, up to now, the helical metallic surface states of Bi-based 3D TIs have not unambiguously demonstrated evidence of unconventional Josephson effects related to Majorana physics in topological Josephson junction devices.

Proximity-induced superconductivity in Josephson junction devices fabricated 
on thin films of Bi$_2$Te$_3$ and other Bi-based TIs has also been reported recently~\cite{WZWu_2021,SCharpentier_2017,MPStehno_2020,PSchuffelgen_2018,MPStehno_2016,LGalletti_2017,Schuffelgen_2019,PSchuffelgen_2019}. Advantages of using thin films over single crystals include their design flexibility that enable the fabrication of several Josephson junction devices alongside numerous Hall bar structures on the same sample. This provides the opportunity to obtain a coherent quantum-transport picture of the sample that includes the Josephson supercurrent transport and the transport carrier densities from magnetotransport, all of which are  extracted from the same thin film. Here, recent experimentally observed induced superconductivity in topological Josephson junctions fabricated on Bi$_2$Te$_3$ thin films are reviewed. Lastly, potential future quantum-transport experimental exploration of Bi$_2$Te$_3$-based film devices is proposed.

To study the proximity effect in the bulk and the contribution of the bottom surface to the supercurrent, we  recently reported proximity-induced superconductivity in Josephson Nb/Bi$_2$Te$_3$/Nb devices fabricated from thin (6 and 15~nm) Bi$_2$Te$_3$ films with small, but finite residual doping, see Fig.~\ref{fig:Fig-09}\textcolor{blue}{(a)}. The films were deposited on (111)-oriented STO substrates, which were used as the back-gate dielectric for combined gate-tunable studies of the Josephson supercurrent and normal state transport~\cite{MPStehno_2020}. Hallmarks of the Josephson effect were observed in these junction devices: namely, a current--voltage characteristic exhibiting a clear Josephson supercurrent, see Fig.~\ref{fig:Fig-09}\textcolor{blue}{(b)},  and a Fraunhofer-like magnetic field modulation of the critical current~\cite{MPStehno_2020}. In the presence of microwave radiation, current plateaus appear in the \textit{I--V} characteristics of the junction with zero differential resistivity, see Fig.~\ref{fig:Fig-09}\textcolor{blue}{(c)}. These plateaus, called Shapiro steps~\cite{SShapiro_1963}, are due to the ac Josephson effect.
A color map of the differential resistance dependence on the microwave power and critical current is shown in Fig.~\ref{fig:Fig-09}\textcolor{blue}{(d)}, demonstrating the evolution of the Shapiro steps. Unlike the Fraunhofer-like magnetic diffraction patterns, Shapiro steps do not depend on the geometry of the junction. Instead, they depend on the current-phase relation. For a pure $\sin (\varphi/2)$ current-phase relationship contributing a $4\pi$-periodic supercurrent~\cite{CBeenakker_2013,SanJose_2012,MHouzet_2013}, an unconventional sequence of even steps, with missing odd steps, is expected. Conversely, conventional $2\pi$-periodic Andreev bound states will result in Shapiro steps at voltages $V=nhf/2e$, where $n$, $h$, $f$, 
and $e$ are the integer multiples of the step index, the Planck constant, the microwave 
frequency, and the elementary charge, respectively. This characteristic voltage is half the step size of the unconventional  $4\pi$-periodic bound state. From the analysis of the \textit{I--V} curves extracted from the differential resistance color-scale maps, see Fig.~\ref{fig:Fig-09}\textcolor{blue}{(d)}, the spacing between Shapiro steps show that both $\textrm{even} - n$ and $\textrm{odd} - n$ are present at the employed microwave frequency, see Fig.~\ref{fig:Fig-09}\textcolor{blue}{(c)}. 
\begin{figure*}[!t]
\centering
\includegraphics[width=1\textwidth]{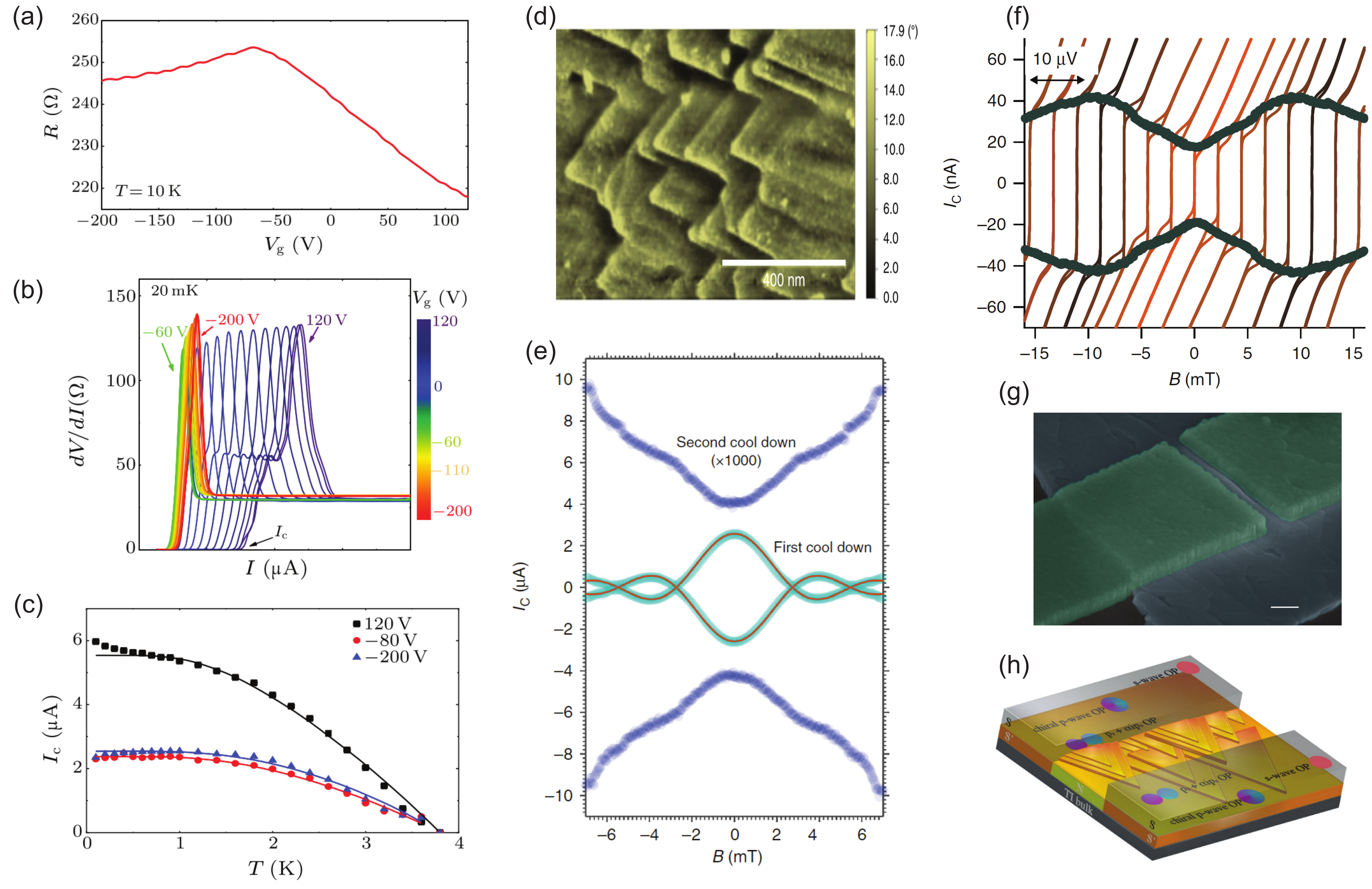}
\caption{\linespread{1.5} (a) Back-gate voltage dependence of junction resistance for Nb$/$Bi$_2$Te$_3/$Nb Josephson junction devices with an 8-nm weak link of Bi$_2$Te$_3$ grown on an STO substrate. The maximum resistance is located at 68~V, considered to be the charge neutrality point. (b)~$dV/dI$ versus $I$ for similar Nb$/$Bi$_2$Te$_3/$Nb Josephson junction devices with a junction length of 134~nm measured at different gate voltages. (c)~Corresponding temperature dependence of the critical current at gate voltages of 120, $-80$ and $-200$~V. (d)~Surface morphology of 80-nm-thick film of Bi$_2$Te$_3$ grown on a GaAs substrate showing a typical growth with aligned pyramidal domains. (e)~Magnetic field pattern of a typical  Al/Bi$_2$Te$_3$/Al junction device at 20~mK, before (cyan) and after (blue points) the thermal cycle.  At the first cool down, a conventional Fraunhofer-like Josephson diffraction pattern is observed, but after the thermal cycle, the critical current is dramatically reduced, and a dip at zero magnetic field appears.  (f)~\textit{I--V} characteristics as a function of magnetic field and temperature for Al/Bi$_2$Te$_3$/Al junction devices. The dark points yield the magnetic pattern of the junction device with a
pronounced dip at zero magnetic field. (g)~Colored SEM image of a typical Al$/$Bi$_2$Te$_3/$Al junction fabricated on a similar sample as in (d)~after the second cool down (white scale bar is 200~nm). (h)~Schematic illustration of a junction device at the surface of a Bi$_2$Te$_3$ film to probe-induced mixed-parity superconductivity, which illustrates the induced order parameter  and the occurrence of $\pi$-paths across the junction.
 The $s$ electrodes (Al) induce a mixed $s+p$-wave superconductivity at the surface of Bi$_2$Te$_3$; only the $p_x+ip_y$-component is shown. (a)--(c)~Reprinted from~\cite{WZWu_2021}, \copyright ~2021 Chinese Physical Society and IOP Publishing Ltd. (d)--(h)~Reprinted from~\cite{SCharpentier_2017}.}
 \label{fig:Fig-10}
\end{figure*}

In the same study, we  also reported a comparative analysis of the gate dependence of both the supercurrent in Josephson junctions and the carrier density from Hall bar structures fabricated side-by-side on the same film. This demonstrated strong correlations between the supercurrent transport and the charge distribution in Bi$_2$Te$_3$ samples~\cite{MPStehno_2020}. By using the substrate as a gate dielectric, the carrier density was tuned electrostatically, and the Josephson supercurrent and the normal-state transport properties were measured in Bi$_2$Te$_3$ thin-film devices. Three gate-voltage ranges with distinct behaviors were identified: one region of intermediate gate bias, where the measured quantities change rapidly with the applied electric field, and two saturation regions for large bias of each polarity,  Fig.~\ref{fig:Fig-09}\textcolor{blue}{(e)}. Similar gate-dependence characteristics were also observed in the carrier density obtained by Hall effect measurements on the same film. Furthermore, excellent agreement was demonstrated between  experimental carrier densities and the calculated carrier distribution in the thin film using a tight-binding model based on the solutions of the self-consistent Schr\"odinger--Poisson equations, Fig.~\ref{fig:Fig-09}\textcolor{blue}{(f)}.
This combined study of the Josephson effect and magnetotransport  clearly demonstrates that the Josephson supercurrent in Nb/Bi$_2$Te$_3$/Nb thin-film devices maps band structure properties of the thin films, thus providing a coherent picture that incorporates band bending in the discussion of proximity effects for 3D TI devices. 

Recently, using the STO substrate as a bottom gate dielectric, Wu \textit{et al.}~\cite{WZWu_2021} also reported qualitatively similar gate-tunable transport measurements on Josephson junctions of Nb/Bi$_2$Te$_3$/Nb devices with  an 8-nm-thick Bi$_2$Te$_3$ film. Both the normal-state resistance and the supercurrent of the thin-film devices were tuned by applying a back-gate voltage in the range of $-200$ to $120$~V, Figs. \ref{fig:Fig-10}\textcolor{blue}{(a)} and \textcolor{blue}{(b)}.
In the $p$-type regime of the TI film, the Josephson current at different applied gate voltages was well described by a short ballistic junction model, Fig. \ref{fig:Fig-10}\textcolor{blue}{(c)}, whereas in the $n$-type regime, diffusive bulk modes emerge for $T < 0.7$~K. The lack of diffusive bulk modes in the $p$-type regime was attributed to the formation of $p$--$n$ junctions~\cite{WZWu_2021}. 

Charpentier \textit{et al.}~\cite{SCharpentier_2017} reported unconventional magnetic field patterns in Al/Bi$_2$Te$_3$/Al Josephson junction devices made using flakes that were exfoliated from Bi$_2$Te$_3$ films grown on (100)-oriented GaAs  substrates. The nanostructured morphology of the flakes exhibited the commonly observed, characteristic triangular-shaped terraces of Bi$_2$Te$_3$ films, see Fig.~\ref{fig:Fig-10}\textcolor{blue}{(d)}. The Josephson junction devices were realized by transferring the flakes onto SiO$_2$/Si substrates~\cite{SCharpentier_2017}. The peculiar result of this work is the observation of a dramatic change in the magnetic field dependence of the Josephson effect  of devices after thermal cycling (from base to room temperature and back to base temperature). The critical current of the junctions was reduced by more than two orders of magnitude, and the magnetic field pattern of
the junction devices was found to have a dip at zero applied
magnetic field. An inverted Josephson magnetic field pattern was observed after thermal cycling, see Figs.~\ref{fig:Fig-10}\textcolor{blue}{(e)} and \textcolor{blue}{(f)}. Strain-related phenomenology that tune the interquintuple layer interactions was invoked to explain this observation. As the exfoliated Bi$_2$Te$_3$ flakes were transferred to SiO$_2$/Si substrates, there is strain during thermal cycling  related to the huge difference between the thermal expansion coefficient of Bi$_2$Te$_3$ and that of the SiO$_2$/Si substrate. Indeed, plastic deformation induced by compressive strain, resulting in buckling of the Bi$_2$Te$_3$ weak link in the nanogap, was observed after a second cool down, Fig.~\ref{fig:Fig-10}\textcolor{blue}{(g)}. 
The observed dip at zero applied magnetic field in the unconventional magnetic field pattern of the Josephson current was presumably attributed to the simultaneous existence of the $0$ and $\pi$ couplings
across the junction provided by a mixed $s + p$-wave
order parameter, where the $\pi$ coupling is due to the combined effect of a sign-changing $p$-component of the order parameter and scattering in the junction devices, Fig.~\ref{fig:Fig-10}\textcolor{blue}{(h)}.

\begin{figure*}[t]
\centering
\includegraphics[width=1\textwidth]{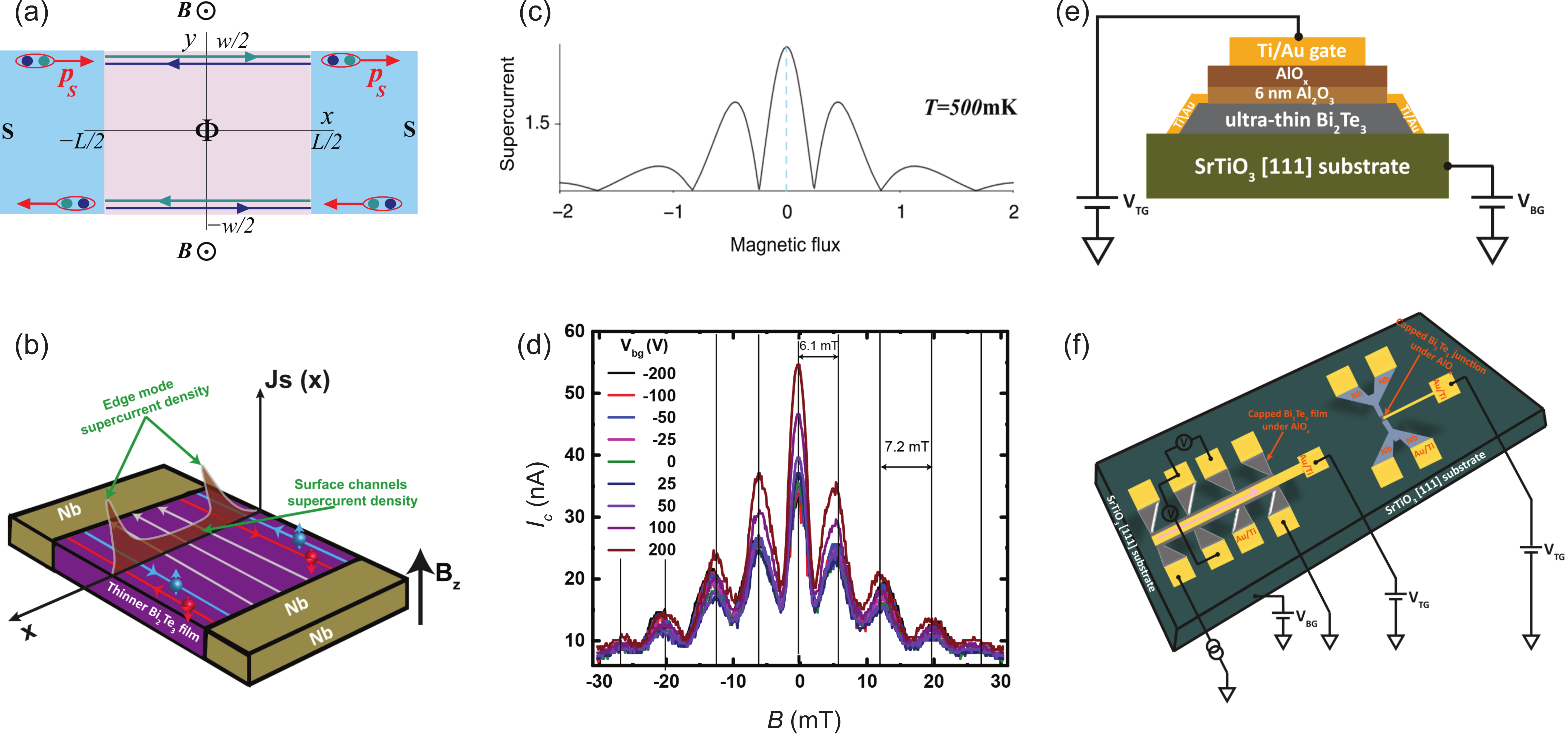}
\caption{\linespread{1.5} Outlook in quantum transport for Bi$_2$Te$_3$-based TI nanostructures.  (a)~Schematic illustration of a 2D TI with helical edge states between two superconductors for out-of-plane applied magnetic field. ~Corresponding  schematic illustration of edge supercurrents in thin Bi$_2$Te$_3$ TI Josephson device with reduced bulk carriers. (c)~Theoretical predicted behavior of supercurrent versus magnetic flux in an edge-dominated supercurrent regime. (d)~Preliminary signature of quantum interference of edge supercurrents in a 6-nm-thick Bi$_2$Te$_3$ TI Josephson device for a film grown on (111)-oriented STO substrate. The device dimensions (width and length) are 860 and 250~nm, respectively. (e)--(f)~Dual-gating configuration in Bi$_2$Te$_3$-based TI nanostructures: (e)~schematics of the envisaged ultra-thin \textit{in situ}-capped Bi$_2$Te$_3$ samples. Using both  top and bottom gates, the sample could potentially be gated towards the QSH regime. (f)~Illustrative image of a TI Josephson device and a Hall bar structure fabricated side-by-side on the same Bi$_2$Te$_3$ thin film together with the measurement configuration. (a)~and (c)~Reprinted with permission from \cite{GTkachov_2015}, \copyright ~2015 American Physical Society.}
 \label{fig:Fig-11}
\end{figure*}

\begin{flushleft} 
\textit{3.5 Prospective Quantum Transport Experiments on Bi$_2$Te$_3$ Thin-Film Devices}
\end{flushleft}
\paragraph*{}
Lastly, for this section on quantum transport, potential future experiments for Josephson junction and Hall bar devices fabricated on Bi$_2$Te$_3$ thin films are discussed. For future investigations,  a promising approach to achieving zero-energy Majorana modes in 3D TIs, particularly in Bi$_2$Te$_3$ TI thin films,  is to use the 1D edge states originating from thin films that are in the QSH regime. A crossover regime from a 3D to a 2D TI phase is anticipated in the band structure of Bi$_2$Te$_3$ due to hybridization effects when the sample thickness is reduced~\cite{CXLiu_2010,YZhang_2010}. In such a 2D TI phase, the sample would be in the 2D QSH regime with the current carried solely by the helical edge states, Fig.~\ref{fig:Fig-11}\textcolor{blue}{(a)}. In this QSH regime, trivial non-perpendicular channels in the junctions, which contribute  gapped $2\pi$-periodic mode Andreev bound states, would be significantly reduced, and the Majorana fermion fingerprints would then appear strongly in ultra-thin Bi$_2$Te$_3$-based devices, Fig.~\ref{fig:Fig-11}\textcolor{blue}{(b)}~\cite{MSnelder_2013}. 

As discussed in theoretical proposals~\cite{GTkachov_2015,JSong_2016}, TI Josephson junction devices (TIJDs) fabricated on such 2D TI samples act as nanoscale superconducting quantum interference devices (SQUIDs) in which a magnetic flux enclosed in the interior of the 2D TI controls the interference of the Josephson currents flowing at the opposite edges of the sample, Fig.~\ref{fig:Fig-11}\textcolor{blue}{(c)}.
Figure~\ref{fig:Fig-11}\textcolor{blue}{(d)} depicts preliminary experimental indications of supercurrent originating from states probably located on the edges of the Josephson junction fabricated on a 6-nm-thick Bi$_2$Te$_3$ film grown on an STO substrate. For all back-gate voltages $V_\textrm{bg}$, the device exhibits unconventional Fraunhofer pattern behavior. As $V_\textrm{bg}$ is decreased, the critical current decreases without changing the overall behavior. The measured shift towards a SQUID-like interference pattern in the 6-nm TIJD suggests a development of peaks in supercurrent density at both edges, coexisting with residual two-dimensional current flow in the junction, Fig.~\ref{fig:Fig-11}\textcolor{blue}{(b)}. These preliminary peculiar $I_\textrm{c}(B)$ characteristics  provide a promising sign towards the formation of edge-mode proximity-induced superconductivity in
thin Bi$_2$Te$_3$ TIJDs in the vicinity of the 2D QSH regime. 

Additional systematic characterization and a complete set of measurements of the Josephson properties of ultra-thin Bi$_2$Te$_3$ TIJDs are needed. The first experimental challenge is to identify and control precisely the thickness range under which the Bi$_2$Te$_3$ thin films are in the 2D  QSH regime. The second challenge is to ensure that the homogeneity of the desired ultra-thin films is very high, so that all the regions of the sample are in the QSH regime, thus avoiding an asymmetry in supercurrent distribution of the subsequently fabricated TIJDs. As discussed in the next section, interest in controlling the growth of Bi$_2$Te$_3$ films in the ultra-thin regime is also motivated by recent theoretical proposals that demonstrate the feasibility of maximizing the TE figure of merit of Bi$_2$Te$_3$ TI films in the few-quintuple layer regime~\cite{JLiang_2016,YXu_2014,NXu_2017}. A third challenge would be to prepare such ultra-thin films on insulating substrates with a high relative dielectric constant, which will be used as a bottom gate in transport experiments. This challenge includes adding an efficient top gate on the top film surface and  capping the film \textit{in situ} before exposing it to ambient conditions to avoid degradation, Fig.~\ref{fig:Fig-11}\textcolor{blue}{(e)}. Hall bar devices and Josephson junction devices should ideally be patterned side-by-side on the same ultra-thin Bi$_2$Te$_3$ sample, Fig.~\ref{fig:Fig-11}\textcolor{blue}{(f)}. The dual-gate configuration will be useful for pushing the sample towards the QSH regime, and then performing systematic studies of both the gate-dependent carrier distributions and Josephson properties in the devices. 

\section*{\Large{4. T\lowercase{opological} s\lowercase{tates for} i\lowercase{mproving the}  T\lowercase{hermoelectric} p\lowercase{roperties of }  u\lowercase{ltra-}t\lowercase{hin} B\lowercase{i$_2$}T\lowercase{e$_3$} n\lowercase{anostructures}}}

Bi-based 3D TIs (e.g., Bi$_2$Te$_3$, Bi$_2$Se$_3$ and Bi$_{1.5}$Sb$_{0.5}$Te$_{1.7}$Se$_{1.3}$) and other TI materials such as Sb$_2$Te$_3$ and SnTe are known  to be good TE materials~\cite{NXu_2017,LMuchler_2013,YXu_2016,ITWitting_2019,TCHsiung_2015}. TE materials directly convert thermal energy into electrical energy and vice versa, and they are considered clean energy converters~\cite{HJGoldsmid_2009}. For practical applications, the conversion efficiency of TE materials is often characterized according to a dimensionless parameter, the TE figure of merit $ZT$ with the expression $ZT=\frac{S^2 \sigma}{\kappa}T$~\cite{HJGoldsmid_2009,ITWitting_2019}, where $S$, $\sigma$, $\kappa$ and  $T$ are the Seebeck coefficient, electrical conductivity, thermal conductivity and absolute temperature, respectively. For detailed reviews of the TE properties of Bi$_2$Te$_3$ on compound classes with both good TE and TI properties, and also discussions of typical TE properties of TIs with gapless boundary states, the interested reader is referred to~\cite{ITWitting_2019,LMuchler_2013,YXu_2016}. 

In recent years, there have been several theoretical and experimental attempts to understand  the contribution of Bi$_2$Te$_3$ TI boundary states to thermoelectricity~\cite{NXu_2017,YXu_2016,YXu_2014,HShi_2015,JChen_2018}. This is because Bi$_2$Te$_3$ is one of the second-generation 3D TIs investigated extensively, and is also the best TE material around room temperature~\cite{ITWitting_2019}. In particular, the focus has been on exploring the possibility of achieving enhanced TE performance in Bi$_2$Te$_3$ TI films that are in the ultra-thin regime (within the thickness range of one to six QLs)~\cite{NXu_2017,JLiang_2016}. Recent progress in  exploring the contribution of TI boundary states to thermoelectricity in  epitaxial Bi$_2$Te$_3$ TI thin films is presented briefly in order to understand the relation between TE properties and topological nature. Special emphasis will be placed on maximizing the TE figure of merit $ZT$ of Bi$_2$Te$_3$ TI nanostructures with topologically protected surface states. 

\begin{figure*}[t]
\centering
\includegraphics[width=1\textwidth]{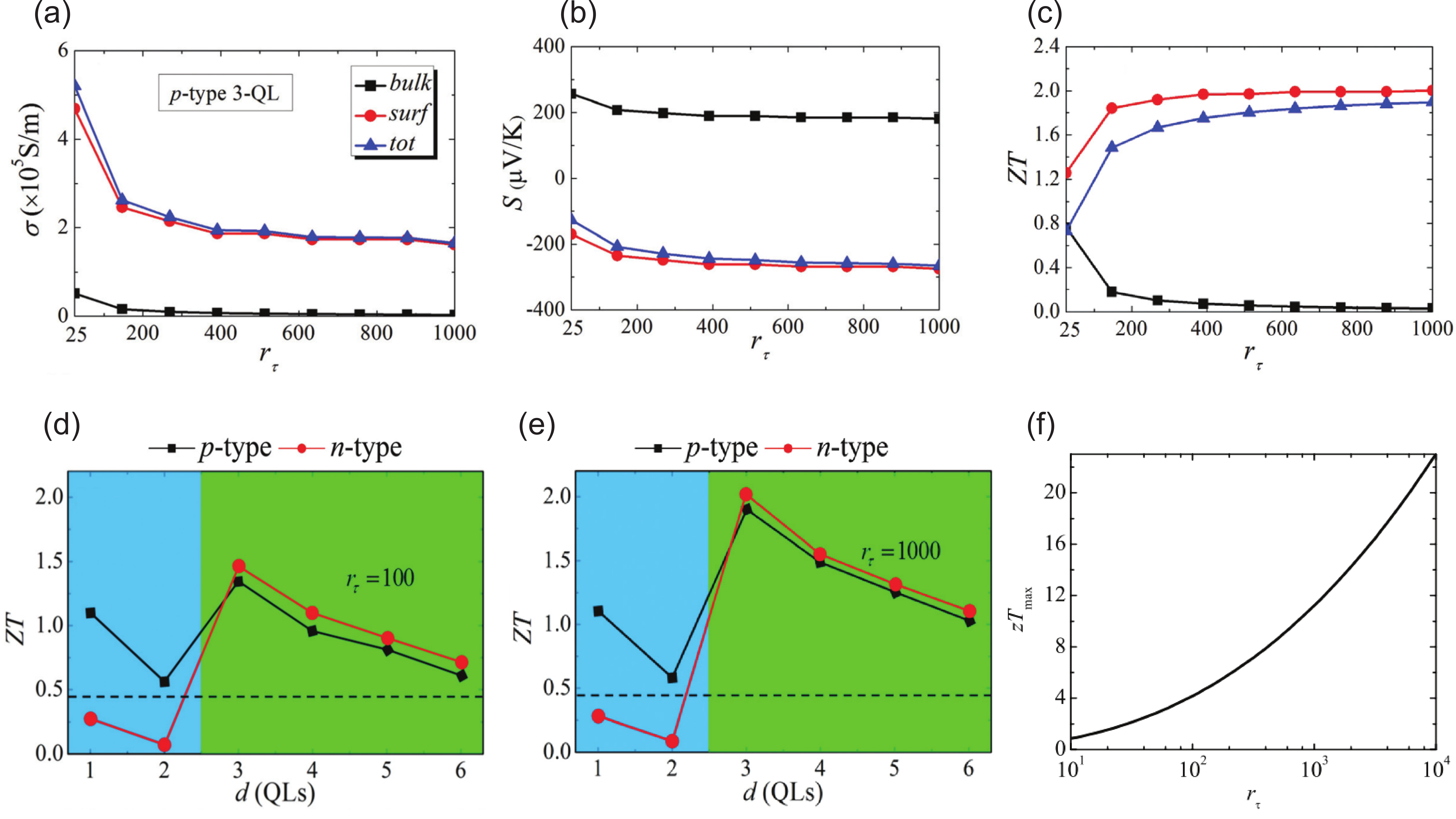}
\caption{\linespread{1.5} Potential of topological states for improving TE properties in ultra-thin Bi$_2$Te$_3$ films. Calculated (a)~electrical conductivity $\sigma$, (b)~Seebeck coefficient \textit{S}, and (c)~figure of merit $ZT$ of the $p$-type three-QL Bi$_2$Te$_3$ film as a function of the relaxation time ratio $r_{\tau}$, where contributions from the surface and bulk states are also determined. (d) and (e)~Calculated thickness-dependent figure of merit $ZT$ values of Bi$_2$Te$_3$ films (blue and green regions indicate films with trivial and non-trivial surface states, respectively), where in the topologically non-trivial region, the relaxation time ratio is assumed to be (d)~100 and (e)~1000. Dashed lines indicate the $ZT$ value of the bulk Bi$_2$Te$_3$ sample in its pristine form. (f)~Maximum possible figure of merit $ZT$ of 2D TIs  as a function of the scattering-time ratio $r_{\tau}$. (a)--(e)~Reprinted with permission from~\cite{JLiang_2016},  \copyright ~2016 The Royal Society of Chemistry. (f)~Reprinted with permission from~\cite{YXu_2014}, \copyright ~2014 American Physical Society.}
 \label{fig:Fig-12}
\end{figure*}

On the theoretical side, using first-principles calculations and the Boltzmann transport theory, Liang \textit{et al.}~\cite{JLiang_2016} reported a systematic exploration of the effects of topologically protected surface states on the TE performance of ultra-thin Bi$_2$Te$_3$ films. By tuning  the relaxation time ratio $r_{\tau}$ between the bulk states and the topological surface states, the maximum $ZT$ value of Bi$_2$Te$_3$ TI films was optimized at a critical thickness of three QLs.  Figures~\ref{fig:Fig-12}\textcolor{blue}{(a)}, \textcolor{blue}{(b)} and \textcolor{blue}{(c)} depict room-temperature transport data (conductivity $\sigma$, Seebeck coefficient $S$) and the $ZT$ value of the $p$-type three-QL Bi$_2$Te$_3$ film as a function of the scattering-time ratio $r_{\tau}$. The surface states dominate the 
electronic transport, Fig.~\ref{fig:Fig-12}\textcolor{blue}{(a)}. The total $ZT$ value is shown to be dominated by the surface states with only minor bulk contributions and increases with increasing  $r_{\tau}$, Fig.~\ref{fig:Fig-12}\textcolor{blue}{(c)}. Similar behaviors were also obtained for the $n$-type three-QL Bi$_2$Te$_3$ film. Notably, the $ZT$ values of the $n$-type system could be enhanced to be as large as those of the $p$-type system~\cite{JLiang_2016}. The $ZT$ values of ultra-thin $p$-type and $n$-type Bi$_2$Te$_3$ films were also  demonstrated to show distinct non-monotonous dependence on the film thickness with a maximum value at three QL when the system enters the non-trivial topological region from the trivial regime, Figs.~\ref{fig:Fig-12}\textcolor{blue}{(d)} and \textcolor{blue}{(e)}. Independently, Xu \textit{et al.}~\cite{YXu_2014} also explored the influence of $r_{\tau}$ on TE performance for a 2D TI by neglecting the lattice thermal conductance. The upper limit of the TE figure of merit $ZT_{\text{max}}$ was found to increase monotonically with increasing $r_{\tau}$, Fig.~\ref{fig:Fig-12}\textcolor{blue}{(f)}. In the same work, the contribution of lattice thermal conductance was also explored by considering a finite mean-free path of phonons, and the mean-free path ratio between phonons and boundary electrons was minimized by introducing defects or disorders into the system, thus improving TE performance. A recent theoretical work also proposed that tuning the Dirac point, for example by introducing defects or changing surface chemistry, is probably an efficient way to retain the enhancement of TE properties from quantum confinement in TI thin films~\cite{MSLim_2018}.

These theoretical works clearly demonstrate the possibility of achieving enhanced TE performance by utilizing TI boundary states in TI films under certain conditions, taking into account the hybridization effects in boundary states. In particular, they demonstrate the feasibility of maximizing the TE figure of merit for Bi$_2$Te$_3$ films with topological surface states in the few-QL regime by increasing the relaxation time ratio between the surface and bulk states and by  introducing disorders and impurities~\cite{JLiang_2016,YXu_2014,NXu_2017}. These theoretically predicted, enhanced $ZT$ values for films that are in the few-QL regime have rarely been reported experimentally for epitaxial Bi$_2$Te$_3$ TI thin films. This is mainly because it is challenging for MBE-grown films to become homogeneous ultra-thin ($\leq 3$ QLs) and high-quality epitaxial TI films, in which all regions of the film are in the same few-quintuple layer regime.   Further experimental efforts are needed to assess the underlying relationship between TE performance and the topological nature of Bi$_2$Te$_3$ thin films in the few-QL regime and  to explore the advantages of TI boundary states for improving Bi$_2$Te$_3$ outstanding TE properties. 

\newpage
In summary, recent progress in epitaxial film growth, magnetotransport, and proximity-induced superconductivity in Bi$_2$Te$_3$ TI films was presented as well as the potential impact of TI boundary states for improving TE properties
of ultra-thin Bi$_2$Te$_3$ films. Using the MBE technique, we succeeded in preparing high-quality, bulk-insulating Bi$_2$Te$_3$ TI films. These films offer the flexibility to control the sample thickness down to a few QLs in the vicinity of the 2D QSH regime, in addition to the possibility of  preparing films \textit{in situ} and capping them immediately after growth to avoid extrinsic defects. Recent quantum transport experiments reported interesting phenomena such as the planar Hall effect and quantum oscillations arising from  high-mobility bulk electrons. In addition,  successful realization of combined gate-tunable Josephson supercurrent and normal-state transport in Bi$_2$Te$_3$ thin-film devices have been achieved. For future quantum transport experiments, it is imperative to use ultra-thin Bi$_2$Te$_3$ films in the 2D QSH regime in order to reach the single-channel regime, which is the ideal experimental configuration to explore  Majorana physics this system. Several theoretical works have discussed the inherent connections between thermoelectric properties and topological states in Bi$_2$Te$_3$ films. However, to explore more the underlying relationship between TE performance and the topological nature, further experimental efforts are needed. In conclusion, clearly tremendous opportunities exist at the intersection of the rapidly evolving fields of TE materials and the new quantum materials of topological insulators. Progress in controlling the epitaxial growth of Bi$_2$Te$_3$ films down to few QLs is expected to reveal the potential of topological states for improving the TE properties of Bi$_2$Te$_3$, which will help  define a path towards   realizing a variety of high-performance TE devices.

\section*{\Large{ A\lowercase{cknowledgments} }}
The author thanks Alexander Brinkman and  colleagues for valuable comments and beneficial discussions, in particular Martin P. Stehno and Hiroaki Myoren, with whom the author collaborated  on the TI project. The author also gratefully acknowledges helpful discussions with Jochen Mannhart and Thomas Whittles, technical support by Yi Wang and editorial support by Lilli Pavka.

\newpage
\section*{\Large{ R\lowercase{eferences} }}

\bibliography{references-2021}
\end{document}